\documentclass[useAMS,usenatbib]{mn2e}
\usepackage{times}
\usepackage{graphicx}

\title[Optical Periodicities of SMC Be/X-ray binaries]{On the periodicities present in the optical light curves of SMC Be/X-ray binaries}

\author[A. J. Bird et al.]{A.J.Bird$^{1}$\thanks{E-mail: ajbird@soton.ac.uk}, M.J.Coe$^{1}$, V.A.McBride$^{1,2,3}$ and A.Udalski$^{4}$
\\
$^{1}$Physics and Astronomy, University of Southampton, SO17 1BJ, UK.\\
$^{2}$Astronomy, Gravity and Cosmology Centre, Department of Astronomy, University of Cape Town, Rondebosch, 7701, South Africa \\
$^{3}$South African Astronomical Observatory, PO Box 9, Observatory, 7935, South Africa \\
$^{4}$Warsaw University Observatory, Al. Ujazdowskie 4, 00-478 Warszawa, Poland 
}
\begin{document}

\date{Accepted 2012 April 23. Received 2012 April 23; in original form 2011 August 19}

\pagerange{\pageref{firstpage}--\pageref{lastpage}} \pubyear{2012}

\maketitle

\label{firstpage}

\begin{abstract}
We present a comprehensive study of the periodic variations observed in OGLE I-band light curves of SMC Be/X-ray binaries, discovering new optical periodicities in 9 systems. We find that these periodicities derive from a number of mechanisms, notably disturbance of the decretion disk on the orbital period of the system, and aliased non-radial pulsations. We develop metrics that allow these mechanisms to be distinguished on the basis of the shape of the folded optical light curve, and use these metrics to categorise the periodicities present in $\sim$50 SMC binary systems. 
We conclude that extreme care must be taken in the interpretation of the  OGLE light curves since only around 30\% of the periodicities present can be unambiguously attributed to orbital periods.

\end{abstract}


\section{Introduction}

Be/X-ray binaries are an important subclass of high-mass X-ray binary (HMXB) that consist of a Be star and a neutron star in a binary system. Mass transfer is believed to occur initially via a semi-stable equatorial decretion disk around the Be star, some fraction of which is captured by the neutron star at, or near, its periastron passage. These episodes of mass transfer result in X-ray outbursts, during which the spin period of the neutron star can often be observed. Knowledge of the orbital period of the system is a key diagnostic tool as it allows the system to be placed on the Corbet diagram of P$_{\rm spin}$ vs P$_{\rm orb}$ \citep{1986MNRAS.220.1047C} that is used to identify the broad classes of HMXB systems. Orbital periods for many Be/X-ray binaries have been determined from period analysis of the X-ray light curves (driven by the spacing of the X-ray outbursts) and Doppler variations in their pulse periods, but are still unknown in many systems. For this reason, ways have been developed to try to use the optical emission as an alternative means to determine the orbital period.

The optical signature of Be/X-ray binaries derives from a number of components: (i) the Be star itself, (ii) the Be star decretion disk, and (iii) any transient accretion disk around the neutron star. The optical light curves of Be/X-ray binaries show considerable differences in overall form, but often show long-term large variations. Similar, if not identical, effects are seen in isolated Be stars \citep{2002A&A...393..887M}, and so these variations are typically interpreted as due to changes in the decretion disk size.

Periodicities, as opposed to aperiodic long-term variations,  in the optical signature are expected from several sources on several timescales:
\begin{enumerate}
\item Radial and non-radial pulsations of the Be star are often present in the 0.1-2.0 day period range.
\item Perturbation of the decretion disk by the orbiting neutron star as it passes periastron, indicating the Be-star orbital period, usually in the range 10--500 days.
\item Semi-stable non-uniformities of the decretion disk, such as disk warping, present on longer (years) timescales.
\end{enumerate}

The on-going OGLE\footnote{\tt{http://ogle.astrouw.edu.pl/ogle3/xrom/xrom.html}}
 coverage of the SMC, which effectively started with OGLE II in 1996, has provided a wealth of optical (I-band) data on the counterparts to the X-ray binary systems. With essentially daily coverage for most of each year, it has been possible to relate the optical to the X-ray data and frequently explain the characteristics of the X-ray behaviour. In addition, regular periodic behavioural patterns in optical data have, in several instances,  been clearly identified with the binary period of the system -- see, for example, \citet{2008MNRAS.384..821M} who showed correlated optical and X-ray outbursts every 138d in SXP46.6. However, \citet{2006AJ....132..971S} have indicated that care must be taken when interpreting all modulations as due to binary periods, since it is possible for non-radial pulsations in the Be star to beat with the daily sampling regime and produce peaks in power spectra that may be confused with binary behaviour. Furthermore, there is evidence for possible super-orbital modulation in many of these systems, a phenomenon well characterised by the behaviour of the supergiant system SMC X-1 (see, for example, \citet{2007ApJ...670..624T}) that may further complicate the study of these systems. An extensive study of such possibilities is presented in \citet{2011MNRAS.413.1600R}.

In this paper, we present a systematic analysis of the OGLE light curves of 49 SMC Be/X-ray binary systems
in which we have searched for both persistent and transient periodic behaviour. We present evidence of both pulsation-related and orbital variations visible in the OGLE light curves from a sample of SMC sources, often transient in nature, and introduce new metrics that allow some differentiation of the various periodic signals in the light curves.

\section{Data Analysis}

\subsection{The OGLE datasets}

The OGLE datasets used in this analysis provide both I-band and V-band photometry. The bulk of this work makes use of the I-band data because it is both much more comprehensive in sampling, and we expect that decretion disk emission contributes a significant fraction of the I-band flux. The V-band light curves have been used as supporting data in some analyses.

\begin{figure*}
\includegraphics[scale=1.05]{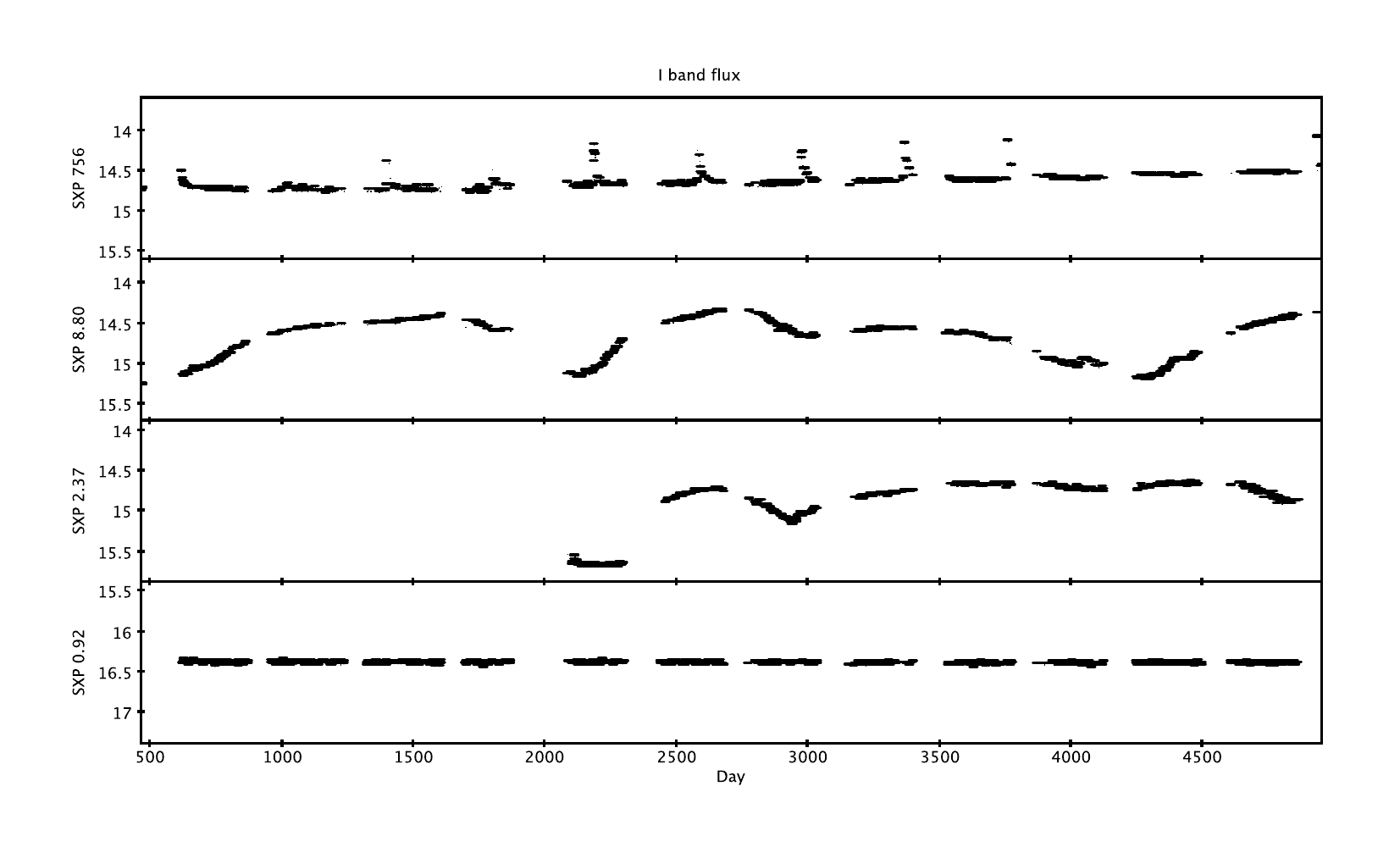}
\caption{Long-term OGLE II and OGLE III I-band light-curves for SMC sources. The light curves shown many different overall morphologies typical of Be systems - see \citet{2002A&A...393..887M} for further examples.}
\label{fig:samples}
\end{figure*}

Some example OGLE I-band light curves are shown in Figure~\ref{fig:samples}, illustrating various examples of optical variability in SMC binary systems. SXP0.92 shows an essentially stable I-band flux through the 12 years of OGLE II and OGLE III monitoring. 1300 I-band observations were taken with a mean measurement error of $\delta$I=9.3mmag; the source varied by only $\Delta$I=85mmag overall on a mean of I=16.39. SXP2.37 was covered only by OGLE III, and so has only $\sim$700 measurement points, but varies by more than one magnitude around a mean of I=14.88. SXP8.80 shows a similar variation, in this case covered by both OGLE II and OGLE III monitoring. SXP756 shows very clear periodic brightening by $\Delta$I$\sim$0.5 over a stable level of I$\sim$14.6 every $\sim$350 days.

Table~\ref{tab:lcparams} presents an overview of the data quality and coverage of the I-band and V-band photometry of the 49 objects studied. For each light curve, we indicate the date range covered, the number of data points, the mean magnitude, the average measurement error ($\delta$) and the long-term magnitude range ($\Delta$). Overall, a typical mean I-band magnitude is I$\sim$15 with a measurement error $\delta$I$\sim$7 mmag on each observation.

\begin{table*}
\begin{tabular}{lcccccccccc}
Source & TJD range & N$_I$ &      I      & $\delta$I & $\Delta$I & TJD range & N$_V$ & V         & $\delta$V & $\Delta$V \\
name   &                      &              & (mag) & (mmag)  & (mmag)    &                      &               & (mag) & (mmag)    & (mmag) \\ \hline
SXP2.37 & 2086.9 - 4868.6 & 707 & 14.89 &  6.2 &  1063 & 3326.6 - 4795.6 & 55 & 14.80 &  4.0 &   221 \\
SXP2.763 & 2086.9 - 4868.6 & 711 & 13.94 &  6.0 &   192 & 3326.6 - 4795.6 & 54 & 14.07 &  4.0 &    57 \\
SXP3.34 &  627.9 - 4953.9 & 837 & 15.57 &  6.2 &   144 &  673.8 - 4954.9 & 109 & 15.63 &  6.0 &   129 \\
SXP6.85 &  627.9 - 4862.6 & 646 & 14.70 &  4.8 &   690 &  665.9 - 3712.6 & 43 & 14.68 &  6.8 &   322 \\
SXP6.88 &  626.9 - 4954.9 & 1000 & 15.37 &  6.1 &   164 &  670.9 - 4954.9 & 99 & 15.56 &  5.3 &   801 \\
SXP7.78 & 2086.9 - 4952.9 & 762 & 14.79 &  5.1 &   196 & 3326.6 - 4953.9 & 89 & 14.92 &  4.0 &    75 \\
SXP7.92 &  621.9 - 4868.6 & 1026 & 13.82 &  4.7 &   647 &  665.9 - 4801.5 & 90 & 13.85 &  5.4 &   285 \\
SXP8.80 &  466.6 - 4952.9 & 1094 & 14.71 &  4.8 &   952 &  466.6 - 4953.9 & 130 & 14.71 &  4.9 &   774 \\
SXP9.13 &  466.0 - 4954.4 & 1077 & 16.30 &  9.1 &   284 &  466.6 - 4954.9 & 99 & 16.54 &  7.5 &   126 \\
SXP11.5 & 2203.6 - 4868.6 & 380 & 15.26 &  6.7 &  1002 & 3654.7 - 4801.5 & 48 & 15.25 &  4.0 &   596 \\
SXP15.3 &  466.6 - 4954.9 & 1075 & 14.47 &  4.7 &   250 &  466.1 - 4954.4 & 93 & 14.68 &  5.2 &   343 \\
SXP18.3 &  466.5 - 4952.9 & 1111 & 15.51 &  6.3 &  1232 &  466.6 - 4953.9 & 135 & 15.64 &  5.3 &   862 \\
SXP22.07 & 2104.9 - 4873.6 & 656 & 14.05 &  6.0 &   137 & 3315.7 - 4797.6 & 49 & 14.21 &  4.0 &    61 \\
SXP25.5 &  621.8 - 4954.9 & 1067 & 15.72 &  7.1 &   372 &  645.9 - 4954.9 & 94 & 15.83 &  6.3 &   400 \\
SXP31.0 & 2104.9 - 4873.6 & 656 & 15.31 &  6.0 &   197 & 3315.7 - 4797.6 & 49 & 15.49 &  4.0 &    28 \\
SXP34.08 & 2086.4 - 4954.4 & 724 & 16.94 & 13.3 &   105 & 3326.6 - 4954.9 & 69 & 16.84 &  7.6 &    48 \\
SXP46.6 & 2086.4 - 4954.4 & 1857 & 14.74 &  5.7 &   511 & 3326.6 - 4954.9 & 148 & 14.80 &  4.0 &   310 \\
SXP59.0 &  626.9 - 4954.9 & 1560 & 15.41 &  6.4 &   657 &  670.9 - 4954.9 & 140 & 15.44 &  5.0 &   365 \\
SXP65.8 &  621.9 - 4862.6 & 589 & 15.73 &  6.8 &   320 &  673.8 - 3379.6 & 39 & 15.77 &  5.9 &   118 \\
SXP74.7 &  466.5 - 4954.9 & 1715 & 16.68 & 11.2 &   179 &  467.1 - 4954.4 & 176 & 16.89 &  7.8 &   167 \\
SXP82.4 &  466.6 - 4952.9 & 1105 & 15.08 &  5.2 &   845 &  466.6 - 4953.9 & 131 & 15.21 &  4.8 &   577 \\
SXP91.1 & 2086.9 - 4865.6 & 713 & 14.78 &  5.3 &   126 & 3326.6 - 4791.6 & 50 & 15.02 &  4.0 &    86 \\
SXP101 & 2086.9 - 4954.9 & 737 & 15.61 &  6.9 &   127 & 3326.6 - 4954.9 & 71 & 15.70 &  4.3 &    67 \\
SXP138 & 2086.9 - 4952.9 & 762 & 16.11 &  7.8 &   294 & 3326.6 - 4953.9 & 89 & 16.18 &  5.2 &   125 \\
SXP140 & 2086.9 - 4954.9 & 724 & 15.52 &  6.6 &   550 & 3326.6 - 4954.9 & 67 & 15.76 &  4.6 &   371 \\
SXP152.1 & 2086.4 - 4954.4 & 718 & 15.43 &  6.3 &   182 & 3326.6 - 4954.9 & 67 & 15.69 &  5.4 &   125 \\
SXP169.3 & 2086.4 - 4865.1 & 712 & 15.38 &  6.0 &   192 & 3326.6 - 4791.6 & 50 & 15.57 &  4.7 &    96 \\
SXP172 &  466.6 - 4954.9 & 1075 & 14.42 &  4.7 &   484 &  466.1 - 4954.4 & 95 & 14.47 &  5.1 &   169 \\
SXP202A &  621.9 - 4954.9 & 1040 & 14.65 &  5.4 &   351 &  665.9 - 4954.9 & 105 & 14.75 &  4.8 &   184 \\
SXP202B &  621.4 - 4868.1 & 911 & 15.34 &  6.0 &   259 &  670.4 - 4795.1 & 88 & 15.52 &  5.5 &   181 \\
SXP214 &  466.5 - 4865.6 & 1060 & 15.22 &  4.2 &   447 &  673.8 - 4954.9 & 109 & 15.63 &  6.0 &   129 \\
SXP264 &  621.8 - 4954.9 & 1067 & 15.90 &  7.6 &   283 &  645.9 - 4954.9 & 94 & 16.09 &  6.5 &   146 \\
SXP280.4 & 2086.9 - 4954.9 & 728 & 15.38 &  6.1 &   244 & 3326.6 - 4954.9 & 69 & 15.56 &  4.8 &   198 \\
SXP293 &  621.9 - 4868.6 & 1037 & 14.57 &  4.7 &   288 &  665.9 - 4795.6 & 93 & 14.87 &  5.1 &   143 \\
SXP304 &  627.4 - 4954.4 & 1052 & 15.58 &  6.1 &    68 &  665.9 - 4954.9 & 102 & 15.75 &  5.9 &    55 \\
SXP323 &  621.3 - 4954.4 & 1060 & 15.24 &  5.6 &   216 &  466.6 - 4954.9 & 97 & 15.42 &  6.0 &   261 \\
SXP327 & 2086.9 - 4952.9 & 762 & 16.75 & 11.6 &   442 & 3326.6 - 4953.9 & 89 & 16.70 &  5.4 &   213 \\
SXP342 & 2086.4 - 4954.4 & 1116 & 15.12 &  6.4 &   967 & 3326.6 - 4954.9 & 116 & 15.16 &  4.1 &   570 \\
SXP348 &  627.9 - 4953.9 & 688 & 15.10 &  5.2 &   963 &  665.9 - 4954.9 & 62 & 15.08 &  5.9 &   629 \\
SXP455A &  627.9 - 4954.9 & 1762 & 15.68 &  6.5 &   718 &  665.4 - 4954.4 & 193 & 15.63 &  5.3 &   338 \\
SXP504 &  621.3 - 4868.1 & 1014 & 14.79 &  5.2 &    83 &  670.9 - 4795.6 & 84 & 15.01 &  5.1 &    69 \\
SXP565 &  628.4 - 4948.4 & 845 & 15.83 &  7.0 &   224 &  665.9 - 4948.9 & 78 & 16.02 &  7.0 &   110 \\
SXP645 &  621.8 - 4868.6 & 1030 & 14.66 &  5.7 &   916 &  670.9 - 4795.6 & 87 & 14.69 &  4.8 &   132 \\
SXP701 &  621.8 - 4868.6 & 1030 & 15.70 &  7.1 &   148 &  670.9 - 4795.6 & 87 & 16.02 &  6.8 &   139 \\
SXP726 &  627.4 - 4953.4 & 1022 & 15.49 &  5.8 &   128 &  673.8 - 4954.9 & 127 & 15.66 &  5.7 &   135 \\
SXP756 &  466.6 - 4954.9 & 1080 & 14.64 &  4.7 &   718 &  466.1 - 4954.4 & 98 & 14.91 &  5.1 &   402 \\
SXP893 &  466.5 - 4954.9 & 1080 & 15.95 &  7.7 &   128 &  466.1 - 4954.4 & 98 & 16.27 &  6.7 &   177 \\
SXP967 & 2104.9 - 4868.6 & 690 & 14.30 &  6.0 &   633 & 3326.6 - 4801.6 & 57 & 14.35 &  4.0 &   147 \\
SXP1323 &  627.4 - 4953.4 & 1021 & 14.57 &  5.7 &   153 &  673.8 - 4954.9 & 127 & 14.65 &  4.7 &   124 \\ \hline
\end{tabular}
\caption{Basic parameters of 49 OGLE light curves for SMC Be/X-ray binaries. N$_I$ is the number of data points in the I-band light curve. I is the mean I-band magnitude. $\delta$I is the mean measurement error on the data points for each light curve. $\Delta$I is the overall range of brightness in the light curve (in magnitudes). TJD = JD - 2440000.5. Equivalent parameters are given for the V-band light curves.}
\label{tab:lcparams}
\end{table*}

\begin{figure}
\includegraphics[scale=0.3]{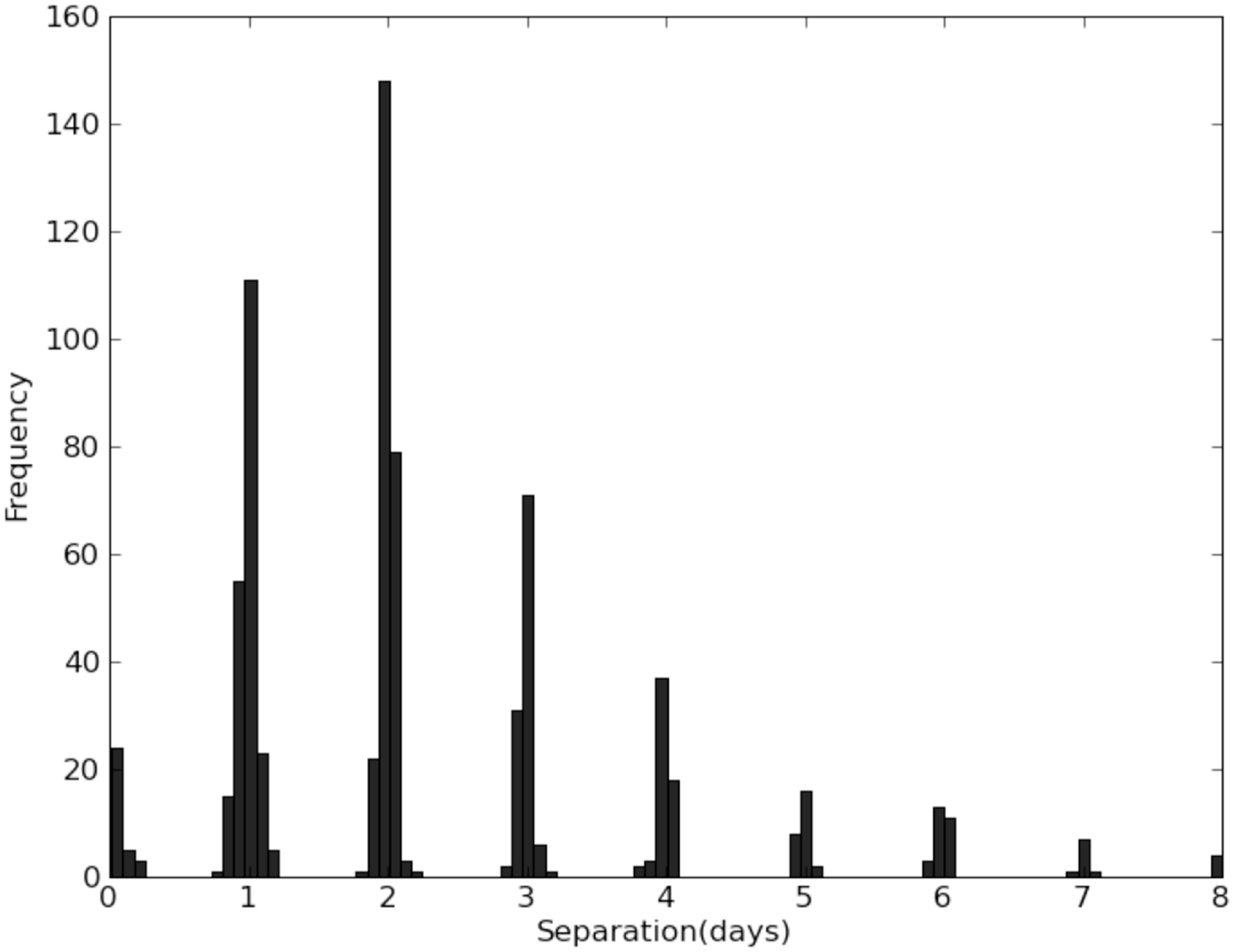}
\caption{Separations of observation times of consecutive observations for SXP327, that may be considered typical of the OGLE light curve sampling.}
\label{fig:sampling}
\end{figure}

Although the observations are optimally at daily intervals, observation limits mean that a typical year of OGLE monitoring produces $\sim$100 usable observations spread over $\sim$220 days, and hence the monitoring periods are separated by $\sim$140 day gaps that are clearly visible in Figure~\ref{fig:samples}. A typical light curve contains $\sim$350 points from OGLE II (if available) and $\sim$800 points from OGLE III. Figure~\ref{fig:sampling} shows the separations of adjacent data points in the light curve of SXP327, which may be considered typical.

\subsection{Analysis method}

We used the fast implementation of the Lomb-Scargle (L-S) periodogram \citep{1989ApJ...338..277P} to search for periodic signals within the dataset, this method generally being preferred for data sets with unequal sampling and data gaps. Periods were searched in the range from 2 days (corresponding to the Nyquist frequency for nominally daily sampled lightcurves) up to 1000 days, after which sensitivity is reduced due to the finite length of the light curves. Significance levels for each periodogram were determined by using Monte-Carlo simulations wherein the data points were randomised in time within the lightcurve, whilst maintaining the same overall time structure. A periodogram was derived for 100000 randomised light curves, and the power level containing 99.99\% of the peak powers was determined. The errors on the derived periods were also derived by Monte-Carlo means, using a {\em bootstrap-with-replacement} simulation performed with many iterations, the error on the period being derived from the distribution of periods recovered from typically 10000 trials.

As the Lomb-Scargle method has limited sensitivity to detect high-frequency modulation in the presence of low-frequency (i.e. long-term) trends in the data, detrending of the light curves was performed to remove long-period signals. A rolling-mean calculation over a specific time window (typically 50--300d), moved in one day steps, was used to determine a detrending model that was then subtracted from the light curve. Such detrending acts as a combined high-pass frequency filter and mean subtraction that should allow the Lomb-Scargle algorithm to work with optimum efficiency when searching for small variations on top of larger long-term variations.

The sensitivity to periodic variations will depend on a number of factors, principally the period and amplitude of the signal being searched for, the length, sampling and noise levels of the light curve being searched, and the ability of the detrending to successfully remove low-frequency noise from the light curve.
We performed a number of simulations to estimate the sensitivity of our analysis to periodic variations. In each case, we created fake light curves with a known periodicity but the same sampling, mean flux, and random measurement noise as real light curves. These fake light curves were then fed through the same analysis process as the true light curves to allow us to assess under what conditions  the simulated periodicity is reliably reconstructed. This procedure was carried out both for 12-year light curves, in order to assess the sensitivity to a persistent periodic signal, and for 1-year light curves, to estimate the sensitivity to transient periodic emission.

We analysed the basic sensitivity to persistent periodic variations at a period of 50d for 4 sample light curves  (SXP6.88, SXP74.7, SXP214 and SXP327) representing different lengths and intrinsic measurement error levels (see Table~\ref{tab:lcparams}). The sensitivity limit for full OGLE II + OGLE III light curves was found to vary between persistent signal amplitudes of 1.5 and 4mmag depending on the light curve quality, and we confirmed that any signal detected above the 99.99\% confidence limit (at a Lomb-scargle power of 20) was reconstructed at the correct period. For one-year datasets, the 99.99\% confidence limit was determined to be at a L-S power level of 14, and the sensitivity to periodicities was found to be closely related to the intrinsic measurement errors -  periodic signals were correctly determined for amplitudes above $\sim$5mmag in SXP6.88 and $\sim$4mmag in SXP214.

\begin{figure}
\includegraphics[scale=0.4]{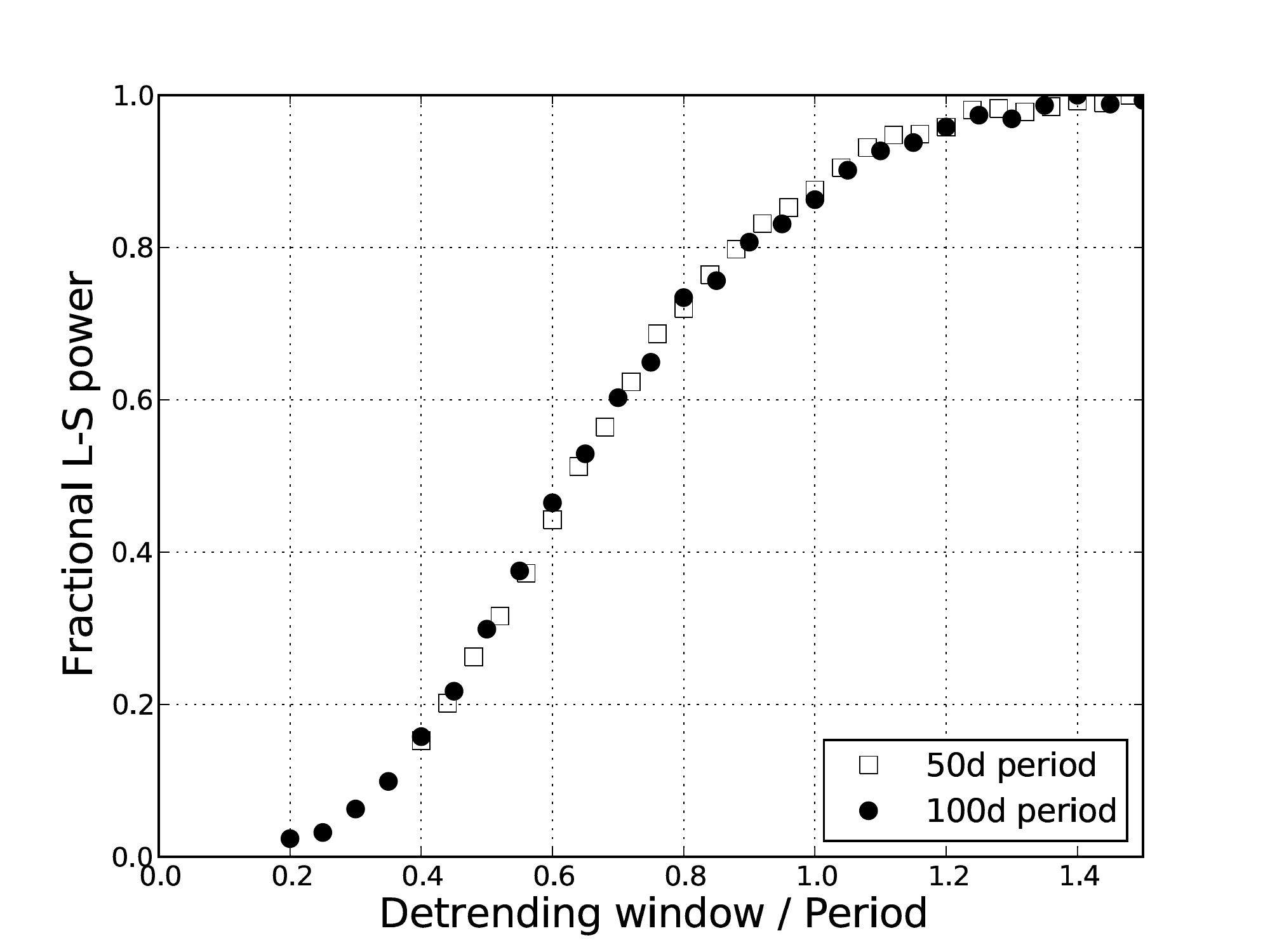}
\caption{Recovered L-S power as a function of detrending window:signal period ratio.}
\label{fig:detrend2}
\end{figure}

We performed additional simulations to quantify the impact of the detrending employed - a series of simulations of a 10mmag periodic signal of either 50-d or 100-d period was injected into a typical lightcurve, which was then detrended with windows of various widths. Only the ratio of the detrending window length to the signal period is important, and the loss of L-S power as a function of this ratio is shown in Figure~\ref{fig:detrend2}). As expected, the L-S power is essentially unaffected when the detrending window is significantly longer than the signal period, but the recovered L-S power is strongly reduced for periods smaller than the detrending window.

Finally, to quantify the effect of the sampling pattern and the length of the dataset on sensitivity, we performed a set of simulations in which we created fake light curves matching the sampling of a typical composite OGLE-II and OGLE-III light curve, added a 10mmag periodic signal of a varying period, and then tested the ability of our algorithms to recover the correct period. The sensitivity was found to be very uniform over a large range of periods, with the exception of a small range around one year periods where the long-term sampling pattern of the data dominated. While the correct periods were recovered at all times, the recovered Lomb-Scargle power, and hence sensitivity, was reduced by $\sim$40\% in narrow windows around one year. This is unlikely to have a significant effect for our studies. Unsurprisingly, it also became difficult to recover the test period when it became comparable with length of the light curve. In the case of yearly subsets of the light curves, periods beyond $\sim$175 days (around $\sim$75\% of the dataset length) could not be reliably detected. Using the combined OGLE-II and OGLE-III light curve lengths, this corresponds to an upper limit on the periods we can detect of $\sim$2000 days, but in practice the detrending windows used made this limit much lower.


\begin{table}
\centering
\begin{tabular}{ccccc} \hline
Source 		& X-ray P$_{orb}$ & Pub P$_{opt}$ & Our P$_{opt}$ & Detrend \\
			&       (days)   &   (days)  &   (days)   & (days) \\ \hline
\hline
SXP 2.37		&	18.38$^{[16]}$ 	&    18.62 $^{[18]}$	&    9.30$\pm$0.01		&      51                   \\
SXP 2.76		&  				&    82.1$^{[2]}$	&   81.81$\pm$0.06 	 	&	101	         \\
SXP 3.34		&				&				&   10.7248$\pm$0.0006		&	 51		\\
SXP 6.85		&     21.9$^{[16]}$	&     114$^{[9]}$	&			&			\\
SXP 7.78		&   44.92$^{[1]}$   	&    44.6$^{[3]}$        	&   44.93$\pm$0.01  	& 	101	         \\
SXP 7.92		&				&    36.8$^{[10]}$ 	&    35.61$\pm$0.05		& 101			\\
SXP 8.80		&   28.5$^{[1,16]}$ 	&				&			& 			\\
SXP 9.13		&    				&	40.17$^{[4]}$	&			&			\\						
SXP 11.5		&   36.3$^{[12]}$ 	&				&			&			\\
SXP 15.3		&      				&    75.1$^{[4]}$ 	&   74.32$\pm$0.03   	&      101		\\
SXP 18.3		&  17.7$^{[1]}$		&    17.7$^{[13]}$ 	& 	17.79$\pm$0.03        	&      101		\\
SXP 22.1		&				&				&   83.7$\pm$0.1		&	101		\\
SXP 25.5		&				&       22.50$^{[20]}$  &   22.53$\pm$0.01    	&	51		\\
SXP 31.0		&				&	90.4$^{[2]}$ 	&   90.53$\pm$0.07		& 	101		\\
SXP 46.6		&   137.36$^{[1]}$	&      137$^{[9]}$ 	&	137.4$\pm$0.2	&	165		\\
SXP 59.0		&   122.25$^{[1]}$	&     60.2$^{[5]}$ 	&			&			\\
SXP 65.8		&         			&	110$^{[14]}$ 	&   111.04$\pm$0.15 	&       201		\\
SXP 74.7		&   33.38$^{[16]}$	&     33.4$^{[4,5]}$ 	&   33.387$\pm$0.006		&	101		\\
SXP 91.1		&    				&     88.2$^{[6]}$ 	&   88.37$\pm$0.03   	&	101		\\
SXP 101		&    				&     21.9$^{[7]}$ 	&   21.949$\pm$0.003   	&	101		\\
SXP 140		&				&     197$^{[8]}$ 	&			&			\\
SXP 169		&      68.5$^{[1]}$	&     67.6$^{[2]}$ 	&    68.37$\pm$0.07		&      101		\\
SXP 172		&	70.4$^{[17]}$	&     69.9$^{[8]}$	&    68.78$\pm$0.08	&        81		\\	
SXP 202B		&				&  229.9$^{[17]}$	&  224.6$\pm$0.3		&     501		\\
SXP 214		&    				&				&    4.5832$\pm$0.0004		&     101		\\
SXP 264		&				&     49.1$^{[4]}$ 	&    49.12$\pm$0.03		&	101		\\
SXP 280		&    				&	127$^{[2]}$ 	&  127.62$\pm$0.25		&	201		\\
SXP 293		&     				&	59.7$^{[6]}$ 	&    59.726$\pm$0.006		&      101		\\
SXP 304		&     				&	520$^{[8]}$ 	&			&			\\	
SXP 323		&     116.6$^{[1]}$	&				&			&			\\
SXP 327		&				&	45.9$^{[15]}$ 	&    45.93$\pm$0.01	&      101			\\
SXP 455		&				&    74.7$^{[6]}$ 	&   74.56$\pm$0.05		&      101		\\
''			&				&      				&   209.6$\pm$0.8		&      301		\\
SXP 504		&    265.3$^{[1]}$	&     273$^{[5]}$ 	&  270.1$\pm$0.5		&      501		\\
SXP 565		&    151.8$^{[1]}$	 &     95.3$^{[6]}$ 	&  152.4$\pm$0.3		&      301		\\
SXP 701		&				&     412$^{[5]}$ 	&			&			\\					
SXP 756		&				&	11.4$^{[19]}$	&	11.404$\pm$0.001	&	31  		 \\	
"			&	389.9$^{[1]}$	&	394$^{[2]}$ 	&	393.6$\pm$1.2 	&	501		\\
SXP 893		&				&				&	3.7434$\pm$0.0005	&	51		\\
SXP 967		&				&	101.94$^{[21]}$			&    101.4$\pm$0.2	&	151		\\
SXP 1323		&				&				&      7.9101$\pm$0.0004		&       25		\\
''			&				&	26.1$^{[8]}$ 	&   26.174$\pm$0.002		&     101		\\ \hline
\end{tabular}
\caption{Sources for which significant periods have been detected. All units are days. References for the published orbital periods are: 
(1) \citet{2008ApJS..177..189G}; 
(2) \citet{2006AJ....132..971S}; 
(3) \citet{2005MNRAS.356..502C}; 
(4) \citet{2005PhDT.........3E}; 
(5) \citet{2005AJ....130.2220S}; 
(6) \citet{2004AJ....127.3388S}; 
(7) \citet{2007MNRAS.376..759M}; 
(8) \citet{2006AJ....132..919S}; 
(9) \citet{2008MNRAS.384..821M}; 
(10) \citet{2009MNRAS.394.2191C}; 
(12) \citet{2009ATel.2202....1T}; 
(13) \citet{2009MNRAS.392..361S}; 
(14) \citet{2007AAS...211.0306S}; 
(15) \citet{2008MNRAS.387..724C}; 
(16) \citet{2011MNRAS.tmp.1085T};  
(17) \citet{2011MNRAS.412..391S}; 
(18) \citet{2008ATel.1670....1S}; 
(19) \citet{2004MNRAS.350..756C}; 
(20) \citet{2011MNRAS.413.1600R};
(21) \citet{2009ATel.1953....1S}
}
\label{tab:SMCperiods}
\end{table}

\section{Period searching}

The detrended light curves were searched for periodic signals with periods between 2 days and 1000 days looking at both the whole light curve and the annual campaigns, subject to the sensitivity limits discussed in the previous section. In the following sections, we report frequencies found in the formally acceptable region from 2 --1000 days, and reserve any comment on the true origin of these periodicities until later sections. Table ~\ref{tab:SMCperiods} indicates the periodic signals detected. We note also the periods previously published in both optical and X-ray light-curves for comparison.

\begin{figure}
\includegraphics[scale=0.3]{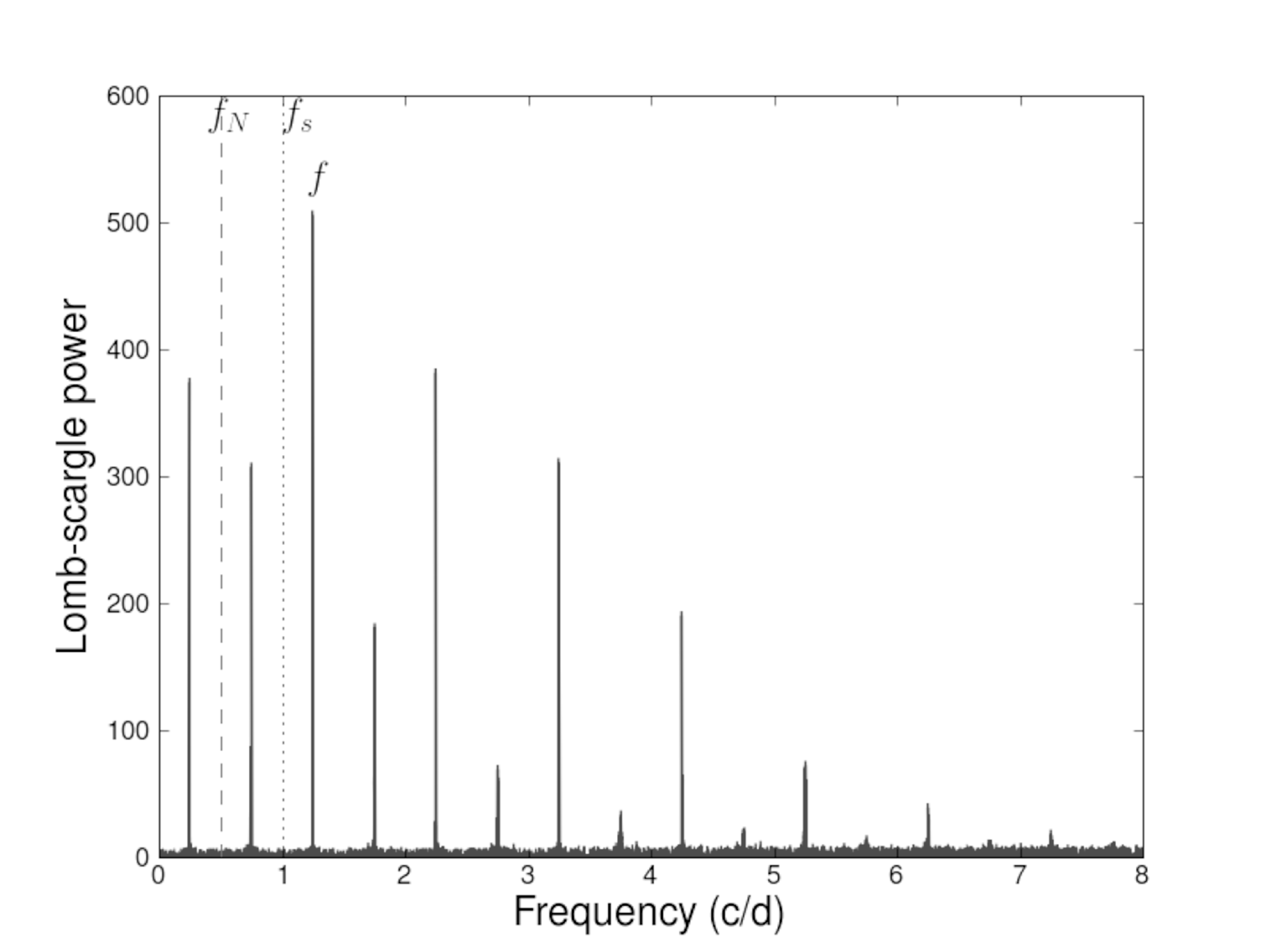}
\caption{Lomb-scargle periodogram for  `fake' sinusoid at period of 0.8 days ($f$=1.25c/d) superimposed on the SXP1323 light curve sampling pattern. The sampling frequency ($f_s$) and Nyquist frequency ($f_N$) are indicated by vertical lines.}
\label{fig:reflections}
\end{figure}

We emphasise here that the recovered period in the 2-1000 day region may not always represent the true period, but may result from aliasing of a periodic signal from outside the formal search range with other periodicities present in the dataset - usually the sampling frequencies of the data. Of particular relevance in the context of this work is the effect of periods around 1 day, that when combined with the sampling pattern associated with the OGLE project datasets (Figure~\ref{fig:sampling}) can produce complex aliasing patterns (often referred to as window functions). Figure~\ref{fig:reflections} shows a L-S periodogram of a simulated sine wave with a period of 0.8 days combined with a typical OGLE sampling pattern. Apart from the fundamental frequency at 1.25 c/d, a complex set of aliases spanning from 0 to $\sim$8 cycles/day, together with the usual reflections about the Nyquist frequencies can all be observed. The broad envelope of the peaks matches, and is clearly driven by, the data sampling shown in Figure~\ref{fig:sampling}.

\begin{figure}
\includegraphics[scale=0.4]{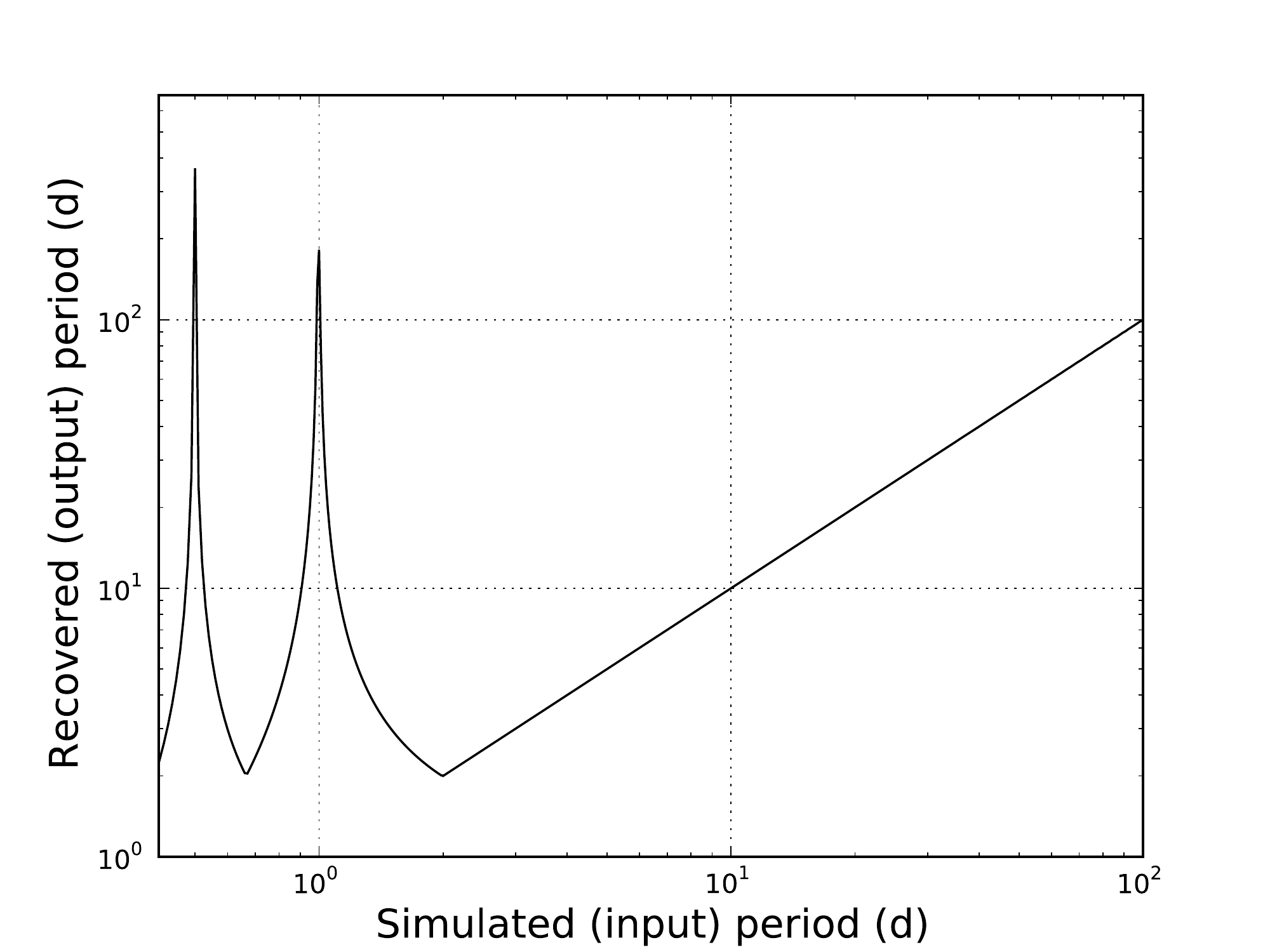}
\caption{Recovered period in the 2 -- 1000 day region for simulated periodicities in the range 0.4 -- 100 days.}
\label{fig:aliasdegen}
\end{figure}

Although the formal Nyquist frequency of the datasets with nominal daily sampling is 0.5 cycles/day (a 2 day period) we cannot ignore the effect of periods below 2 days when searching. Due to aliasing, periods of very near to 1 day can inject power into the periodogram far from the Nyquist frequency, and into the typical range for Be/X-ray orbital period searching. For example, a period of 0.98d corresponds to 1.02 c/d, creating an alias at 0.02 c/d that could easily be misinterpreted as a 50d period. This problem is illustrated in Figure~\ref{fig:aliasdegen}, which shows the detected period (in the formally useful region between 2 and 100 days) resulting from superposition of periodic signals of various periods onto a typical OGLE-II and OGLE-III sampling pattern. There is a clear degeneracy in the solution as a recovered period could result from a number of true periods, and some {\em a priori} knowledge of the expected periods, or some other discriminator is required to break this degeneracy. A possible discriminator, the shape of the folded light curve, is discussed in the following section. One further implication of this pattern is for drifting periods (see Section~\ref{sec:drifters}) - while the alias of a drifting period will still manifest itself as a drifting period of $>$2 days, neither the period or rate of change of the original drifting signal can be uniquely determined as a result of this degeneracy.

\section{Folded light curve profiles}
\label{sec:folding}

Where a clear periodicity was detected in a light curve, or annual section of a light curve, we created a folded light curve for inspection. It was noted that the folded light curves fell into two broad categories and were either sinusoidal variations, or sharply peaked variations often with `fast rise exponential decay (FRED)' shaped profiles.

In order to quantify the shapes of the folded light curves, we defined two metrics for the folded light curves. The first is the full-width half-maximum in phase units, hereafter the phase span (PS). For a sinusoidal variation, the expected value for this metric is 0.5; values below 0.5 would be expected for sharper peaks. The second measures the asymmetry of the peak and is defined as the ratio between the distances {\em in phase space} between the peak and points at 10\% of maximum on the right and left sides of the peak, hereafter the phase asymmetry (PA). For a sinusoid, with a symmetric peak, the expected value is 1; values greater than one indicate a peak with a faster rise than fall.

The folded light curve analysis was performed for those sources exhibiting strong persistent or quasi-persistent (seen in more than one year of data) periodicities. The resulting PS and PA metrics are listed in Table~\ref{tab:PAPS}, together with information on the subsets of data they have been extracted from.

\begin{table*}
\centering
\begin{tabular}{ccccccccc} \hline
Source 		& Date range   		& Period	& Match 	& DW  	& PS 	& PA		& Amplitude	& Amplitude \\		
	 		& (TJD)  			& (d)		&  		& (d)  	& 	 	& 	 	&	($\times10^{-5}$ Jy)		&      (mmag)	\\ \hline		
SXP 2.37		& 4191 -- 4549		& 18.60	& Y	         & 51.0 	& 0.24       & 3.00       	& 16.0	& 37.9 \\
SXP 2.763	& all				& 81.8	& ?		& 101	& 0.35	& 1.38	& 14.1	& 16.9 \\
SXP 3.34		& all				& 10.72	& ?		& 51		& 0.52	& 0.85	& 7.5		& 38.2 \\
SXP 7.78		& all				& 44.93	& Y		& 101	& 0.3		& 1.5		& 14.5	& 38.4 \\
SXP 7.92		& all				& 35.61	& ?		& 101	& 0.63	& 1.33	& 24.0	& 23.5 \\
SXP 15.3		& all				& 74.34	& ?		& 101	& 0.53	& 0.75	& 7.2		& 15.1\\
SXP 18.3		& 3101 -- 3824		& 17.79	& Y		& 51		& 0.33	& 0.86	& 21.8	& 116.0 \\
SXP 22.1		& 3101 -- 5000		& 84.9	& ?		& 121	& 0.47	& 0.88	& 11.0	& 15.0 \\
SXP 25.5		& all				& 22.53     & ?		& 51		& 0.27	& 1.50      	& 3.6		& 21.6 \\
SXP 31.0		& all				& 90.53	& ?		& 101	& 0.33	& 5.5		& 8.8 	& 39.4\\
SXP 46.6		& all excl. 3100--3800	& 137.38	& Y	& 165	& 0.13	& 2.50	& 10.9	& 25.7\\
SXP 65.8	    	& all				& 111.04	& ?		& 201	& 0.33	& 2.67	& 5.1		& 29.4\\
SXP 74.7		& all				& 33.38	& Y		& 101	& 0.33	& 1.0		& 1.9		& 30.7\\
SXP 91.1		& all				& 88.37	& ?		& 101	& 0.33	& 1.1		& 20.0 	& 58.6\\	
SXP 101		& all				& 21.95	& ?		& 101	& 0.40	& 1.29	& 5.9		& 31.9\\	
SXP 169.3	& all				& 68.37	& Y		& 101	& 0.24	& 4.0		& 7.2		& 34.4\\
SXP 172		& all				& 68.785	& Y		& 81		& 0.46	& 2.33	& 8.1		& 14.0\\
SXP 202B		& all				& 224.57	& Y		& 501	& 0.23	& 2.7		& 6.3		& 28.8\\
SXP 214		& all				& 4.58	& ? 		& 101	& 0.5		& 0.6		& 13.0	& 67.0\\
SXP 264		& all				& 49.11	& ?		& 101	& 0.2		& 2.5		& 4.7		& 36.3\\
SXP 280.4	& all				& 127.6	& ?		& 201	& 0.37	& 2.33	& 7.0		& 33.4\\
SXP 293		& all				& 59.72	& ? 		& 101	& 0.37	& 1.63	& 41.4	& 108.0\\
SXP 327		& all				& 45.9	& ?		& 101	& 0.23	& 2.4		& 14.0	& 205.8\\
SXP 455		& all				& 74.55	& ? 		& 101	& 0.3		& 0.88	& - & -\\
			& all				& 209.7	& ? 		& 301	& 0.65	& 1.11	& - & - \\
SXP 504		& all				& 270.1	& Y 		& 501	& 0.4		& 3.6		& 4.4		& 12.5\\
SXP 565		& all				& 152.4	& Y		& 301	& 0.2		& 1.29	& 3.4		& 24.6\\
SXP 756		& all 				& 393.6	& Y		& 501	& 0.1		& 6.33	& 80.0	& 228.5\\
			& non-outburst		& 11.404  & N		& 31		& 0.57	& 1.37	& 9.7		& 25.3\\
SXP 893		& 1283 -- 3101		& 3.74       & ?            & 51.0       & 0.59       & 1.0        	& 2.9		& 26.5\\
SXP 967		& 2200 -- 4700		& 101.37	& ?		& 151	& 0.33	& 2.17	& 22.0	& 34.4\\
SXP 1323		& all				& 7.91	& ?		& 25		& 0.55	& 0.67	&16.0	& 32.9\\
			& all				& 26.17	& ?		& 101	& 0.43	& 1.08	& 24.0	& 49.8\\
\hline
\end{tabular}
\caption{Folded light curve parameters for periodic sources in the SMC sample.
The match parameter indicates when the X-ray and optical periods match thus confirming an orbital period, while a ? indicates that the X-ray period is unknown.  Columns indicate the detrending window (DW), the phase span (PS) and the phase asymmetry (PA) from folded light curve analysis -- see text for more details.}
\label{tab:PAPS}
\end{table*}

\begin{figure}
\includegraphics[scale=0.4]{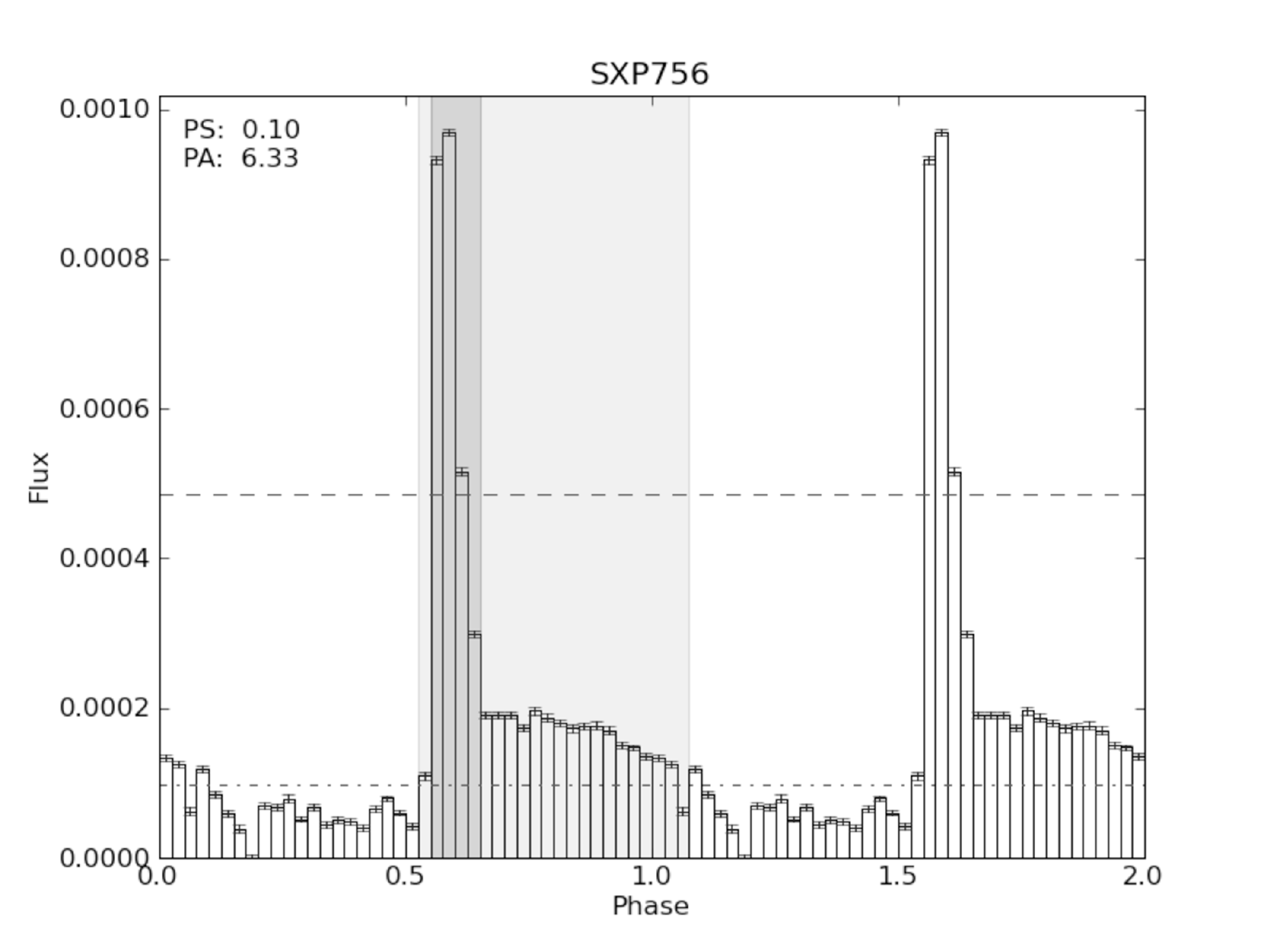}
\caption{Folded light curve analysis for SXP 756 (folded on the 393.6d period). The dashed horizontal lines indicate 10\% and 50\% of the maximum value, where the phase span (darker shaded region) and phase asymmetry (lighter shaded region) are evaluated.}
\label{fig:folded756}
\end{figure}

Two folded light curves are shown to illustrate the principle, representing rather extreme cases. The light curve for SXP756 (Figure~\ref{fig:folded756}) when folded on the orbital period shows very sharp peaks occupying a narrow phase range, and showing a large asymmetry. The darker shaded region indicates the measured phase span (here PS=0.1) and the lighter shaded region indicates the region used in calculation of the asymmetry (here PA=6.33 indicates a very asymmetric peak).

\begin{figure}
\includegraphics[scale=0.4]{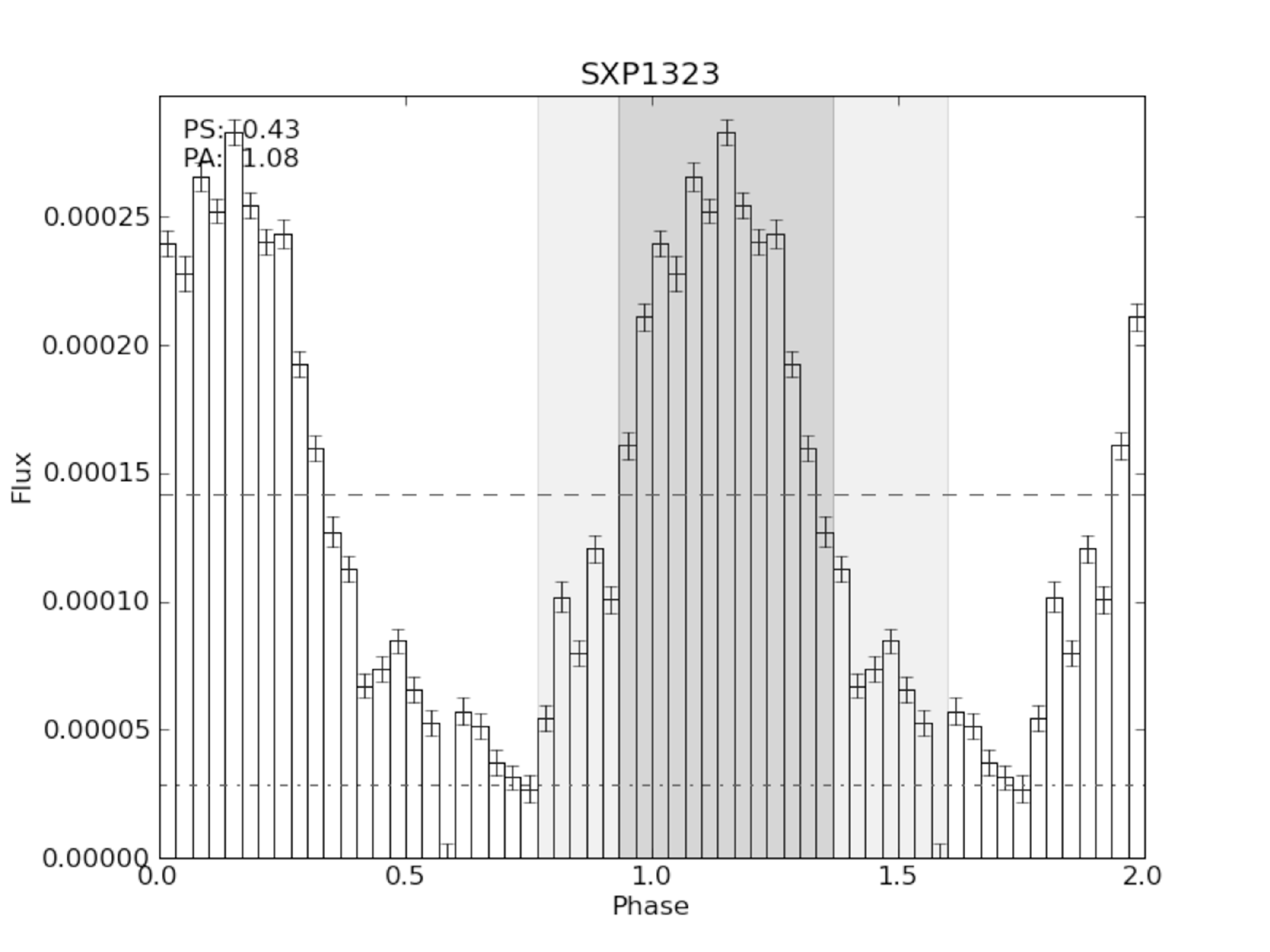}
\caption{Folded light curve analysis for SXP1323 (folded on the 26.17d period). The dashed horizontal lines indicate 10\% and 50\% of the maximum value, where the phase span (darker shaded region) and phase asymmetry (lighter shaded region) are evaluated.}
\label{fig:folded1323b}
\end{figure}

The folded light curve for SXP1323 folded on one of the two detected periods (Figure~\ref{fig:folded1323b}) shows much more sinusoidal behaviour with PS=0.43 and PA=1.08.

\begin{figure*}
\includegraphics[scale=0.55]{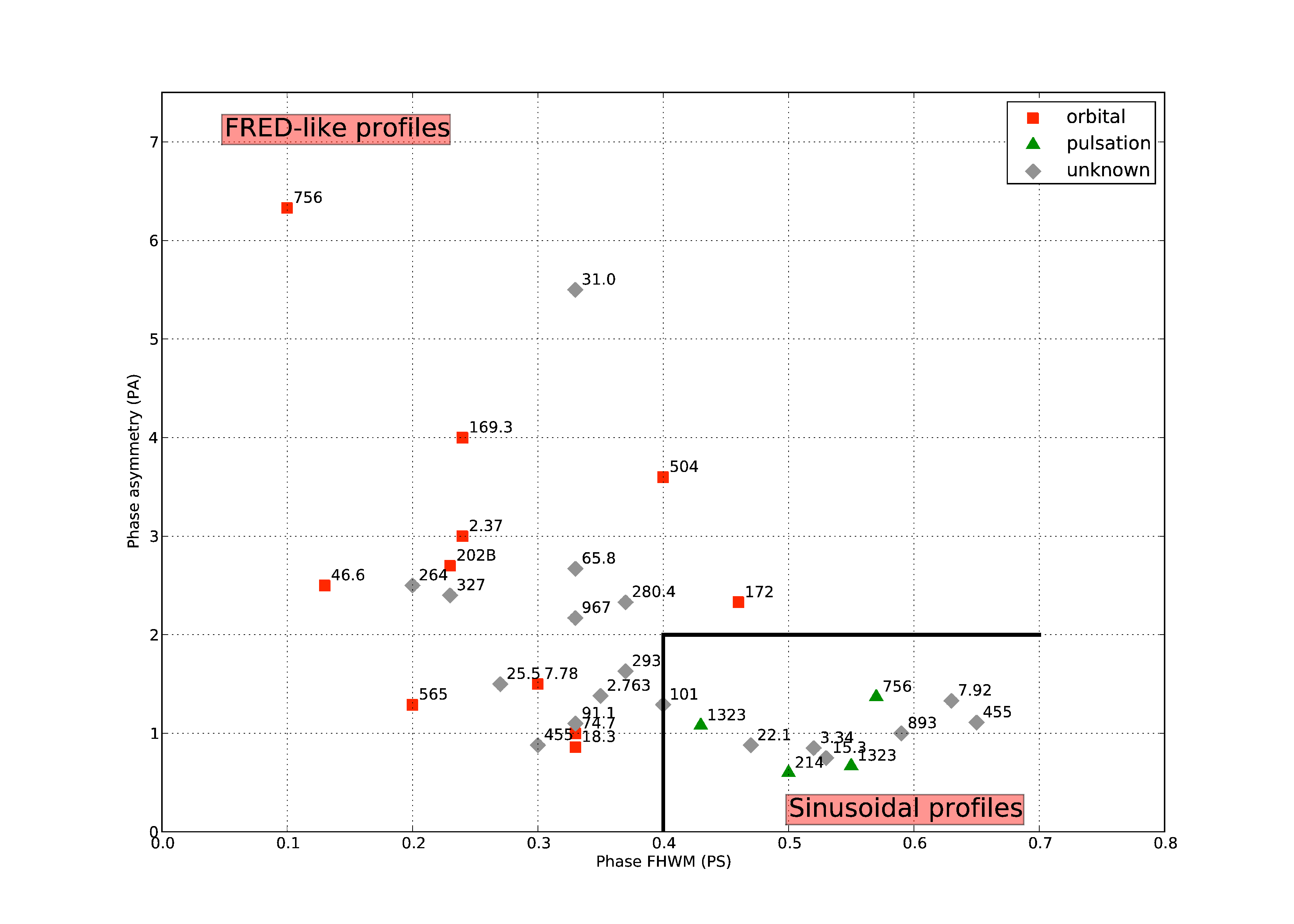}
\caption{Folded light curve metrics for the SMC Be/X-ray binary sample. The solid lines indicate the approximate boundary between regions occupied by sinusoidal (likely aliased pulsations) and FRED-like (likely orbital) light curve profiles.}
\label{fig:freddiness}
\end{figure*}

The folded light curve parameters for those SMC binaries for which folded light curve analysis was possible are shown in Figure~\ref{fig:freddiness}. It is apparent that while some of the SMC systems are clustered around (PS,PA)=(0.5,1.0) point where sinusoids are expected to lie, others show a broad range of folded light curve parameters.

\section{Discussion}

The ``Match'' column in Table~\ref{tab:PAPS} indicates those sources for which the period detected in optical monitoring is also seen in long-term X-ray light curves. In Be/X-ray systems, accretion and the resulting X-ray emission normally only occurs when the neutron star is close to the Be star decretion disk, and there is no strong evidence for stable accretion disks forming around the neutron star. Hence, since the X-ray period can only realistically be due to orbital motion, we can use this to select a subset of sources where we believe the {\em optical} period is indicative of the orbital motion. These sources are SXP18.3, SXP46.4, SXP169.3, SXP172, SXP504, SXP565 and SXP756. There is a strong possibility that SXP280.4 also falls into this category, as the X-ray and optically derived periods are a factor of 2 different. Where the folded light curve can be parameterised, the majority of these sources fall far from the (0.5,1.0) point in Figure~\ref{fig:freddiness} and so are clearly associated with FRED-like profiles.

Conversely, those sources where there is evidence that the optical period is of non-orbital origin cluster around the (0.5,1.0) point. For example, SXP1323 has two strong periods, which is difficult to explain as orbital motion. Furthermore, the two observed periods are strongly incompatible with being orbital based on the Corbet P$_{orb}$-P$_{spin}$ relation for a source with P$_{spin}$=1323s \citep{1986MNRAS.220.1047C}. Similarly, the 11.4d period in SXP756 cannot be orbital, as the 393d orbital period is established beyond doubt. 

We therefore conclude that this `FRED-iness' analysis and diagram is a useful tool to identify optical periods that are likely to be associated with orbital perturbations of a decretion disc based on the shape of the folded optical light curve. Both phase span and asymmetry of the outbursts from disk disruption may be, to some extent, dependent on several orbital parameters - notably period and eccentricity. If we make the naive assumption that the decretion disks in Be/X-ray binary systems all recover with a similar timescale, the phase span parameter will only be effective when the disc recovery time is significantly less than the orbital period. Intuitively, more eccentric systems will show a wider range of interactions between neutron star and decretion disk, while a completely circular system would give a much reduced optical modulation (due only to radial velocity shifts of the spectrum through a stationary filter). Overall, the method should be most sensitive for long-period, eccentric systems, but even reasonably low-eccentricity systems should provide a recognizable orbital signature in the optical light curve. In summary, then, a FRED-like folded optical light curve profile is a strong indicator that the period is orbital. On the other hand, a very sinusoidal folded light curve profile is often associated with non-orbital periodicities, but could still be indicative of orbital motion in a very low-eccentricity system.

\subsection{New orbit determinations}

We have detected previously undiscovered periodicities in the OGLE II and III light curves for several sources, and confirmed previous optical modulations in many others. In some of these cases, the folded light curve analysis gives a strong indication that the periodicity is derived from the orbit of the system, allowing us to propose them as orbital period determinations.

\begin{figure*}
\includegraphics[scale=0.75]{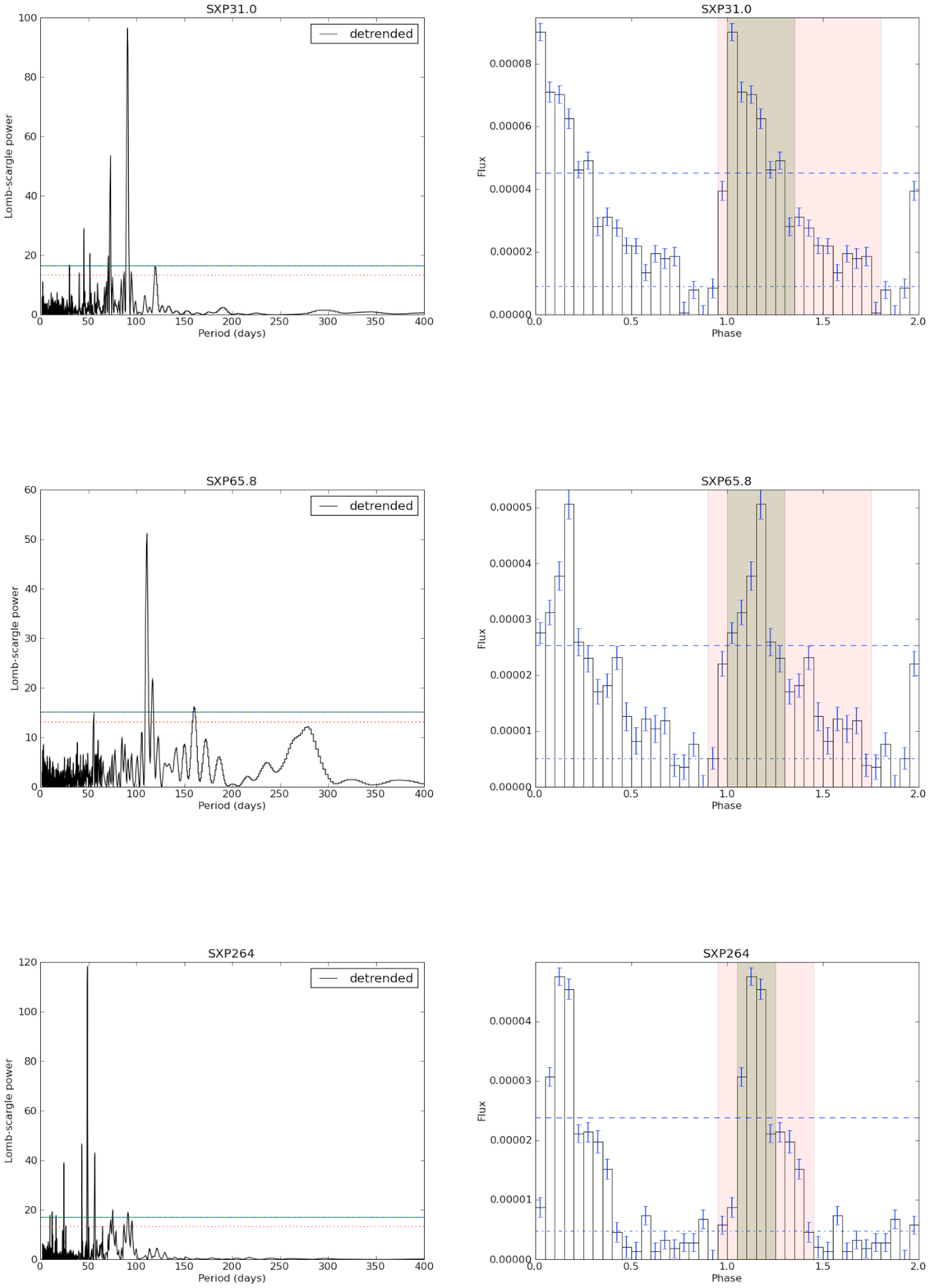}
\caption{Periodiograms (left) and folded light curve analyses (right) for SXP31, SXP65.8 and SXP264, confirming previous optical periods, and strongly suggesting orbital origins for the detected periods. Dashed horizontal lines on periodograms indicate 99 and 99.9\% confidence limits. Dashed horizontal lines on folded light curves indicate 10\% and 50\% of the maximum value, where PS and PA are evaluated (see text for more details).}
\label{fig:orbits1}
\end{figure*}

\begin{figure*}
\includegraphics[scale=0.75]{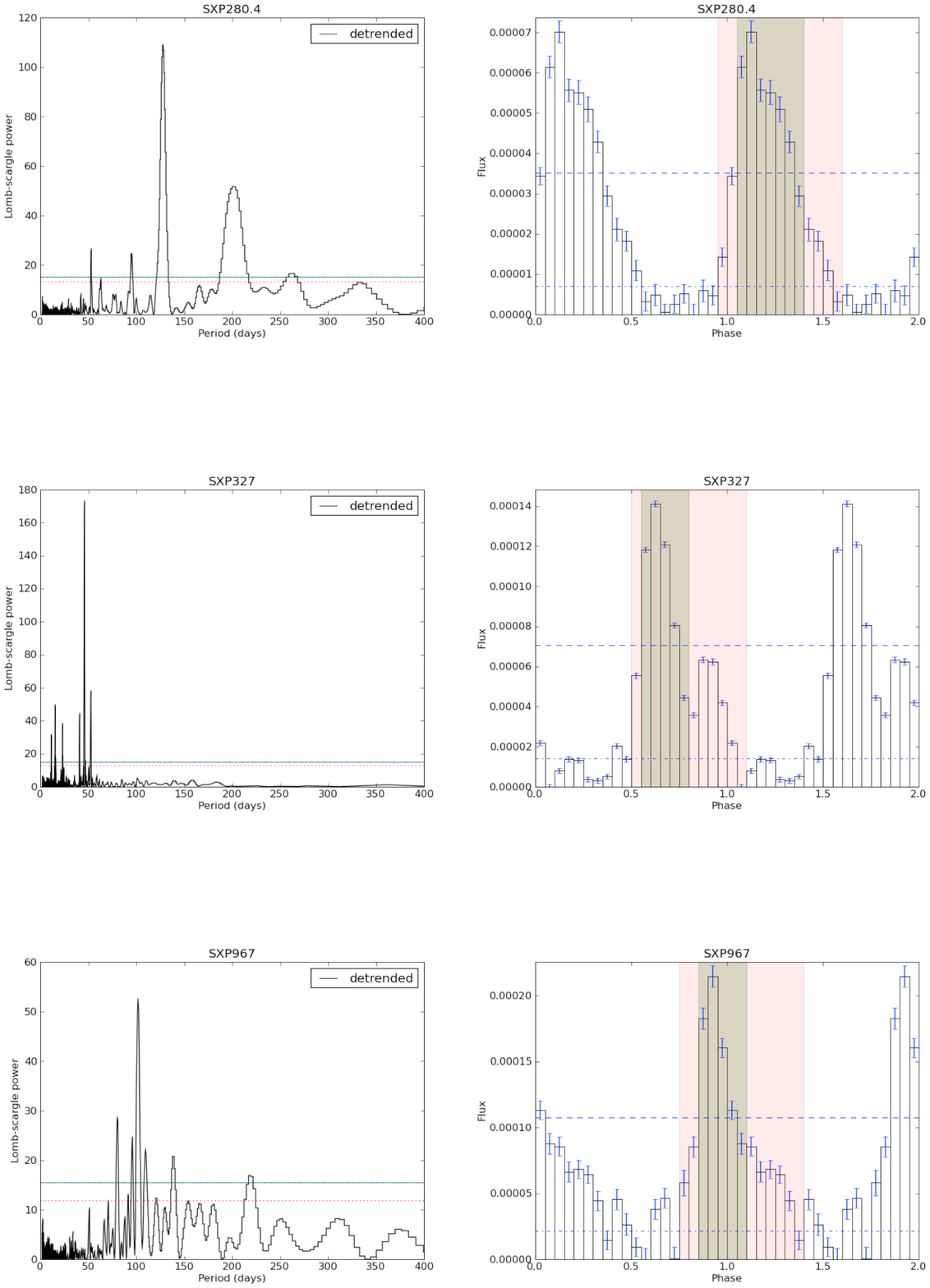}
\caption{Periodiograms and folded light curve analyses for SXP280.4, SXP327 and SXP967, confirming previous optical periods, and strongly suggesting orbital origins for the detected periods. Dashed horizontal lines on periodograms indicate 99 and 99.9\% confidence limits. Dashed horizontal lines on folded light curves indicate 10\% and 50\% of the maximum value, where PS and PA are evaluated (see text for more details).}
\label{fig:orbits2}
\end{figure*}

Specifically, the folded light curve analyses for SXP31.0, SXP65.8, SXP264 and SXP327 provide strong evidence that the measured periods at 90.53, 111.04, 49.11 and 45.9 days respectively, are indeed the orbital periods of these systems. Furthermore, the folded light curves of SXP280.4 and SXP967 show quite convincing evidence of non-sinusoidal (i.e. orbital) modulation at periods of 127.6 and 101.4 days respectively (see Figures~\ref{fig:orbits1} and \ref{fig:orbits2}). The latter is a confirmation of the detection in \cite{2009ATel.1953....1S}, but we note that this is a complex lightcurve, difficult to detrend, and the orbital outbursts are only visible in a subset of the data.

\begin{figure*}
\includegraphics[scale=0.75]{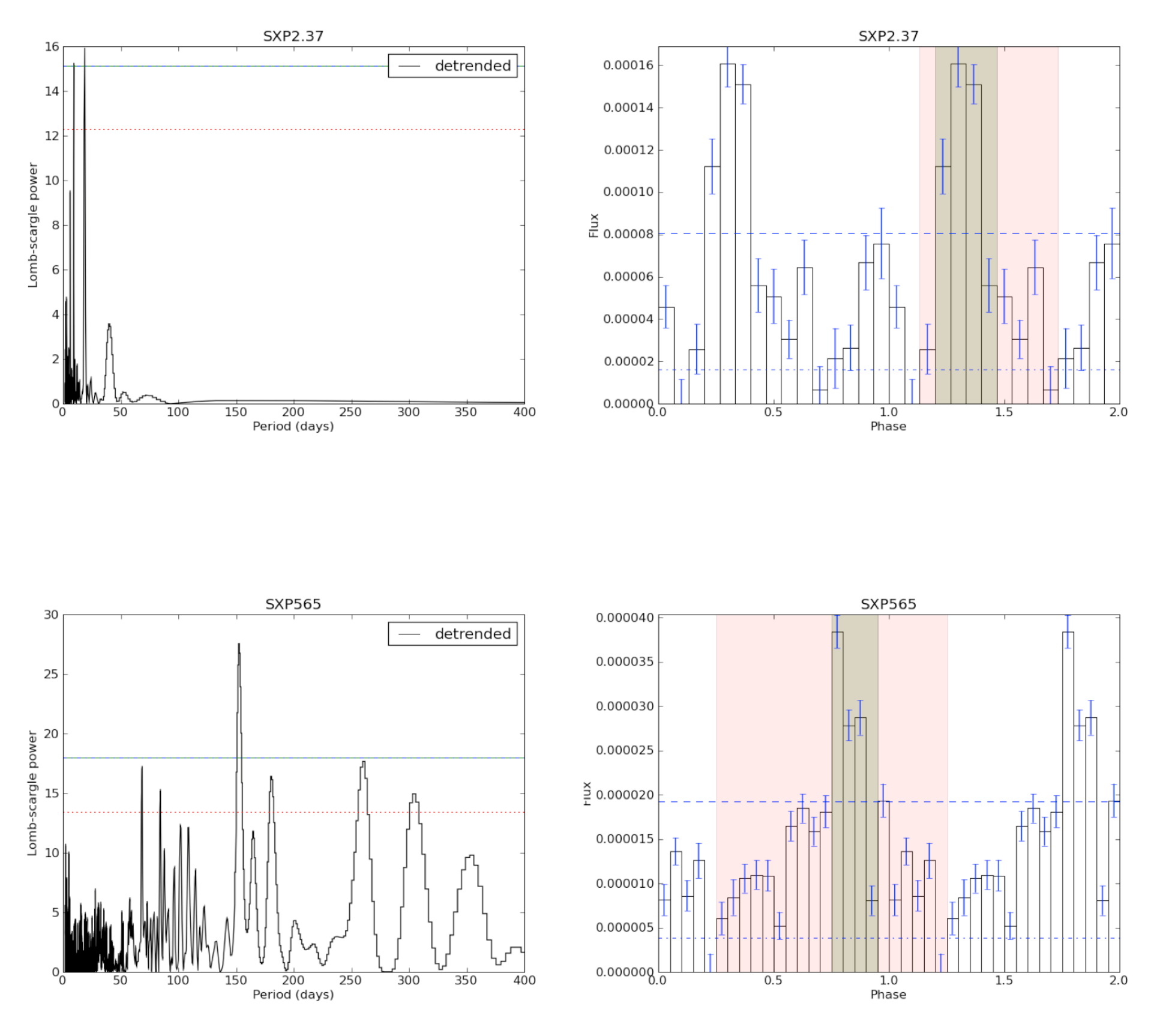}
\caption{Periodiograms and folded light curve analyses for SXP2.37 and SXP565. The first indicates the previously optical period at 18.6 days (now confirmed in the X-ray band) is of orbital origin, the second is the first optical detection of the previously detected X-ray period at 152.4 days. Dashed horizontal lines on periodograms indicate 99 and 99.9\% confidence limits. Dashed horizontal lines on folded light curves indicate 10\% and 50\% of the maximum value, where PS and PA are evaluated (see text for more details).}
\label{fig:orbits3}
\end{figure*}

SXP 2.37 (SMC X-2) is a particularly interesting case, which has been the cause of considerable speculation as to the origin of the periodicities seen in the light curve. Previous studies \citep{2008ATel.1670....1S} have seen periodicities at  $\sim$ 6.2, 9.3 and 18.6 days, which could be interpreted as an orbital period of 18.62d with harmonics. However, this interpretation was challenged by \cite{2009ATel.1992....1S} who interpreted the signals as two aliased pulsations at 0.8592d and 0.9008d with the 18.62d period deriving from the beat frequency between these two pulsations. However, our analysis (Figure~\ref{fig:orbits3}) shows that the transient 18.6d period is firmly in the `orbital' region of Figure~\ref{fig:freddiness}, and we thus conclude that this is by far the simplest explanation for the three periodicities seen in the periodogram. It has also now been shown independently from the X-ray data \citep{2011MNRAS.tmp.1085T} that fitting of the orbital Doppler shifts of the pulsar timing information during a long outburst yields an estimate of the orbital period of 18.38$\pm$0.02d.

For the first time, we confirm that the 152.4d orbital period of SXP565, determined from X-ray measurements \citep{2008ApJS..177..189G}, can also be seen in the optical light curve (Figure~\ref{fig:orbits3}).

\subsection{Sinusoidal variations}

\begin{figure*}
\includegraphics[scale=0.75]{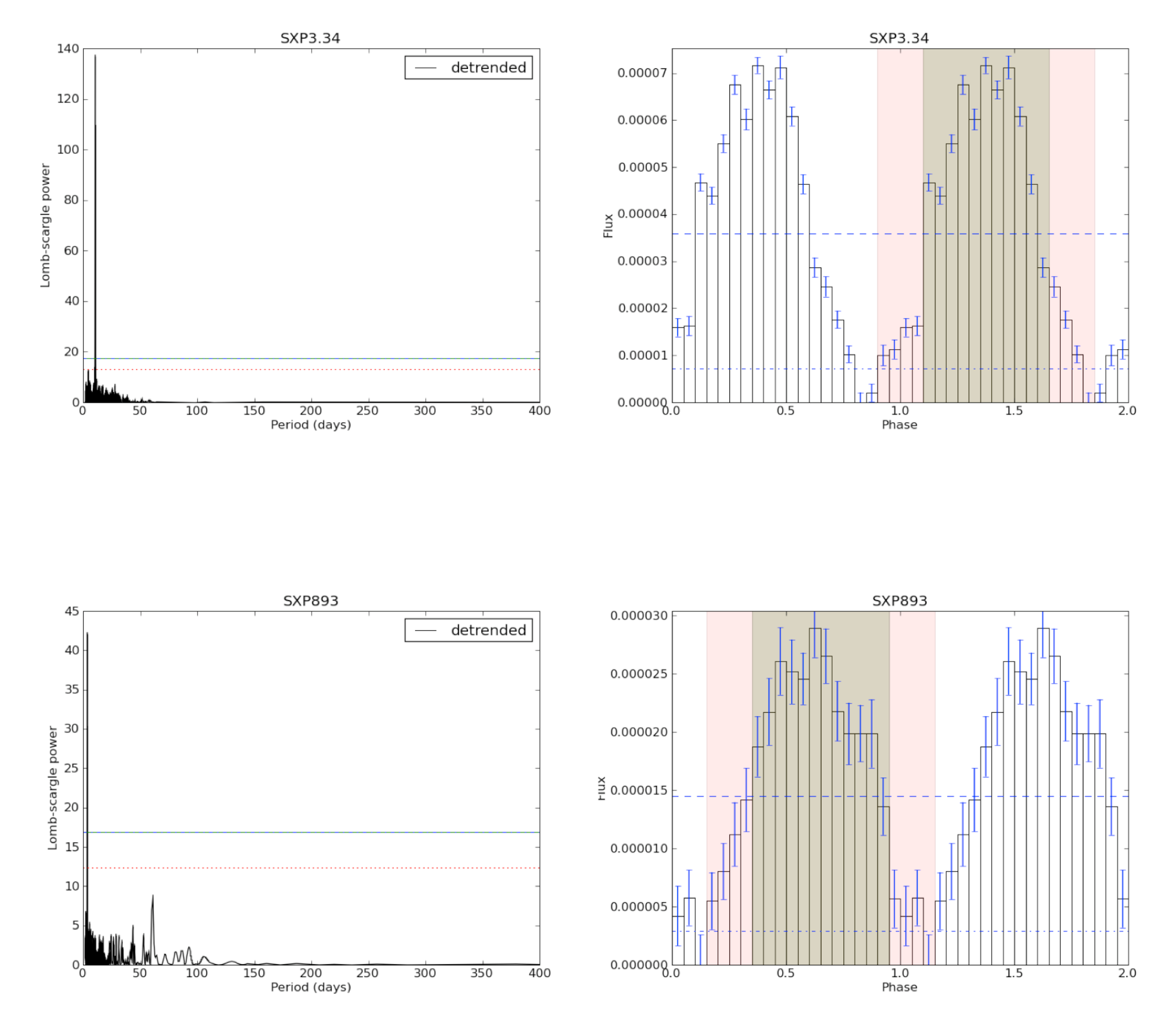}
\caption{Periodiograms and folded light curve analyses for previously undetected signals in SXP3.34 and SXP893. Based on the folded light curve analysis, these two periodicities are unlikely to be due to orbital motion. Dashed horizontal lines on periodograms indicate 99 and 99.9\% confidence limits. Dashed horizontal lines on folded light curves indicate 10\% and 50\% of the maximum value, where PS and PA are evaluated (see text for more details).}
\label{fig:sinus1}
\end{figure*}

The periodicities observed in SXP3.34, SXP7.92, SXP15.3, SXP455 (209.7d period), SXP893 and SXP1323 do not show the characteristics of orbital modulation of a decretion disk, but instead have folded light curves that are essentially sinusoidal in shape. Those in SXP3.34 and SXP893 (Figure~\ref{fig:sinus1}) are newly discovered periodicities, but for obvious reasons we do not propose them as new orbital periods. In particular, the period for SXP893, at 3.74 days, is incompatible with the usual placement of Be/X-ray binaries in the P$_{orb}$-P$_{spin}$ plane.

SXP1323 is an informative case as it shows two strong and persistent pulsations which manifest themselves as 7.91 and 26.17d periods in the OGLE light curves. These periods have been attributed \citep{2006AJ....132..919S} to the aliasing of non-radial pulsations at periods of 0.96 and 0.88 days in the Be star with the 1d sampling of the dataset. Thus {\em apparent} periods at $\sim$8 and $\sim$26 days, with sinusoidal folded profiles, result from non-radial pulsations of the Be star in this system, and this is a plausible explanation for at least some of the other non-orbital periods determined.

As a further example, the second period in SXP756, of 11.4d, is only visible outside of the sharp orbit-induced increases in brightness accompanying the Type I X-ray outbursts that are clearly separated by the 393.6d orbital period. This secondary period has also been attributed to pulsations of the Be star or an orbiting disturbance in the outer decretion disk, and furthermore is seen to change slightly with time \citep{2003AJ....126.2949C,2005MNRAS.356..502C}
  
\subsection{SXP264 - a Rosetta Stone?}

SXP264 is a strange but revealing object because, like SXP756, the optical light curve exhibits periods in addition to a well-established orbital modulation. Previous studies \citep{2005PhDT.........3E} have indicated that it shows a $\sim$49d orbital period in optical data. Our time-resolved analysis, however, shows a slightly different and more complex story. While the expected orbital period is stably detected for the first 10 years of OGLE-II and OGLE-III monitoring, the peak in the time-resolved periodogram shifts significantly in the final 2 years of OGLE-III measurements.

\begin{figure*}
\includegraphics[scale=0.6]{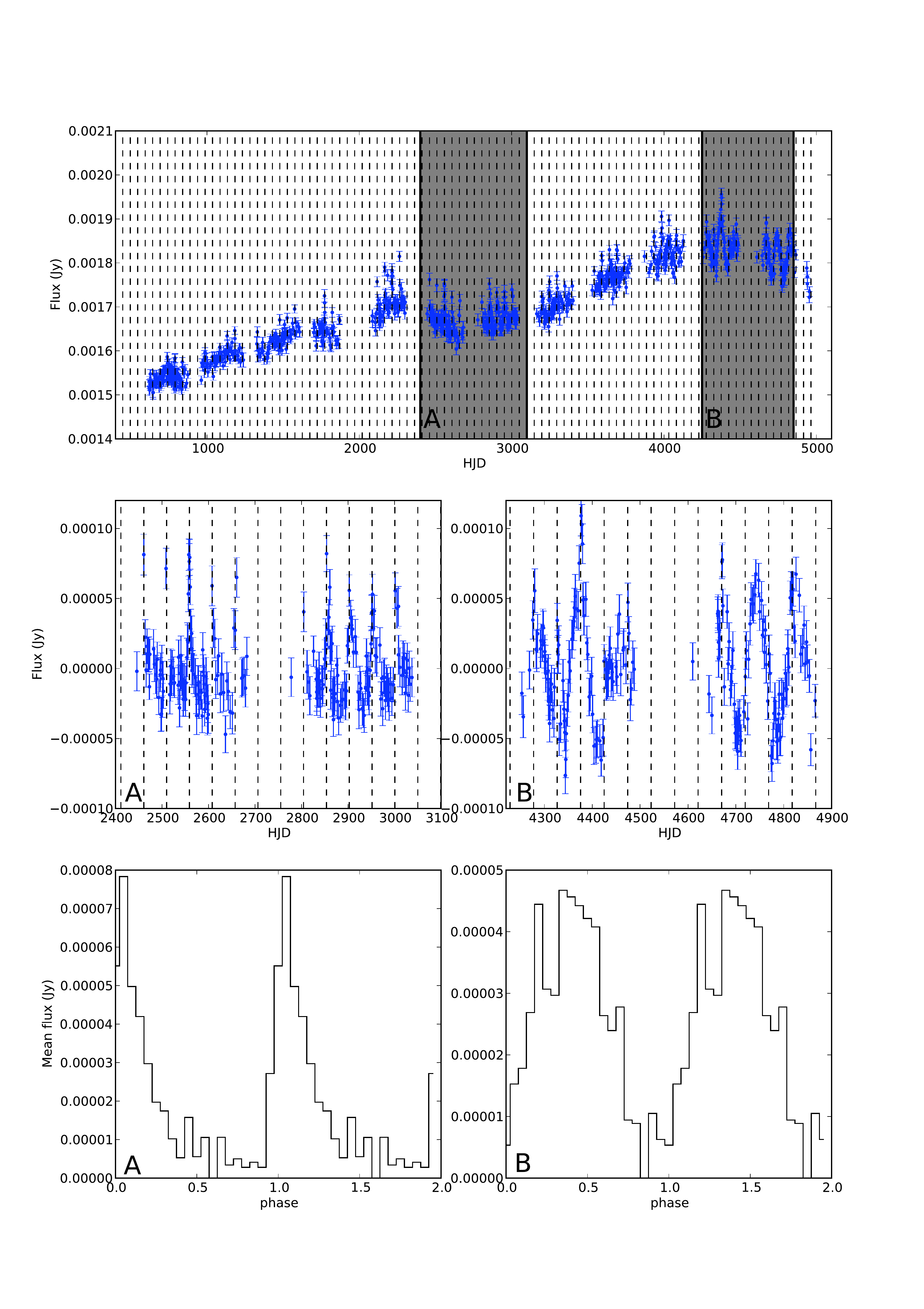}
\caption{The changing shape of SXP264. Two subsections of the light curve show (A) clear orbital modulation on the 49.1day period with a FRED-like folded shape, and (B) a strong sinusoidal modulation on a longer period, with the orbital modulation superimposed. The top panel shows the full OGLE II and III light curve, the middle panels are zoomed views of sections A and B, and the lower panels are the folded light curves - for panel A (lower), the data are folded on a 49.1day period, while for panel (B) lower, they are folded on an 80 day period. Dashed vertical lines in the top three panels indicate expected periastron passages based on knowledge of the orbital period.}
\label{fig:SXP264all}
\end{figure*}

A deeper analysis is summarised in Figure~\ref{fig:SXP264all}. The first 10 years of data yield a folded light curve that shows a clear FRED-like orbital profile at the X-ray confirmed orbital period, as illustrated by a zoom on two years of data in the left centre panel. Using the analysis methods of Section~\ref{sec:folding}, the folded light curve profile is parameterised by PS=0.2 and PA=2.5 which is entirely consistent with a FRED-like orbital modulation. The light curve for the final 2 years however shows a much more sinusuoidal profile with a similar amplitude to that seen in the previous years, but with a visibly different period and shape (right centre panel of Figure~\ref{fig:SXP264all}). While it is hard to be certain with just a few cycles of the $\sim$80-90d oscillations, the period of these sinusoidal variations may not be constant through the two years. 


We interpret these changes in the light curve as a clear indication of both orbital-derived periodicities and non-orbital periodicities, the latter possibly deriving from aliased stellar pulsations, being visible in the same light curve, just as in SXP756. The presence of the sinusoidal variations in the last two years of OGLE III monitoring make it more difficult to automatically detect the orbital variations, but visual inspection of the light curve of the last 2 years around each expected periastron passage (shown by the dashed vertical lines in Figure~\ref{fig:SXP264all}) finds clear increments in the flux at those times, showing that the orbital variations indeed continue. If the non-orbital periodicities at $\sim$80-90d period do derive from aliased pulsations, we can only state that their true period is below 2 days; the degeneracy in the period reconstruction shown in Figure~\ref{fig:aliasdegen} prevents us from unambiguously estimating the true period. There is no indication that the pulsation-derived and orbit-derived variations are mutually exclusive, as they co-exist in this light curve, but the fact that the two sections of the light curve show such clear changes indicates both the transient nature of these phenomena, and the effectiveness of light curve shape as a way of distinguishing them.



\subsection{Transient phenomena}

Several of the sources discussed here display modulations that are transient in nature. In the case of orbital modulations, this is usually modelled by the growth and decline of the decretion disk around the Be star: when the disk is large, the orbital modulation is more pronounced. SXP 18.3 is a good example - no periods may be detected when considering the entire light curve, but the orbital modulations are clearly detectable in a section encompassing approximately two years of monitoring. Other examples may be less extreme but almost all the sources show some variability in the amplitude of the orbital modulation.

\begin{figure*}
\includegraphics[scale=0.33]{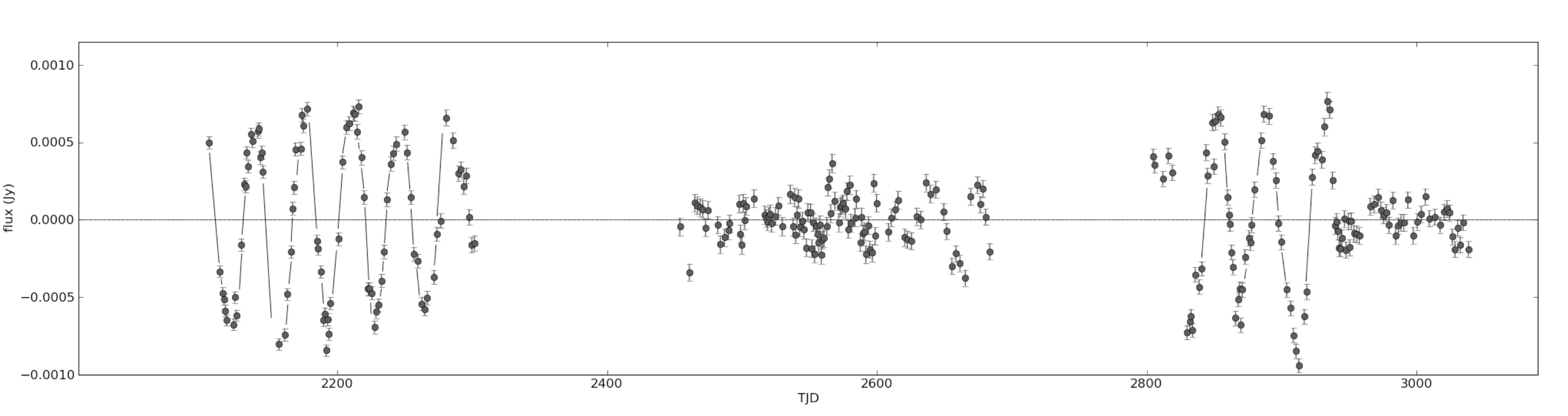}
\caption{Variations in the amplitude of the sinusoidal modulation of the I-band light curve of SXP7.92.}
\label{fig:pulse_variability}
\end{figure*}

The sinusoidal modulations in systems, that we believe to derive from aliased stellar pulsations, are also often seen to be transient in nature. While many are seen at all times throughout the OGLE II and III datasets with almost constant amplitude, others are seen to come and go, and even change in frequency (see next section). Moreover, we sometimes see evidence of very abrupt changes in the pulsation-like behaviour - a classic example being given by SXP7.92 in Figure~\ref{fig:pulse_variability} which shows three consecutive years of OGLE III data. The year-to-year amplitude shows significant changes, but even more dramatic is the abrupt cessation of pulsations at TJD$\simeq$2940.

It is also appropriate to point out here a limitation of the methods used - the Lomb-Scargle method is insensitive to modulations superimposed on large non-periodic changes in brightness of the underlying star. We have attempted to overcome this as far as possible by the use of detrending on appropriate timescales in order to restore a level baseline against which the Lomb-Scargle method works best. However in some systems, rapid underlying changes can dramatically reduce the sensitivity of the period searching for some parts of the dataset. Thus it is important to separate out transient phenomena intrinsic to the source itself and those introduced by changes in sensitivity to modulations. 

\subsection{Period drifters \label{sec:drifters}}

One feature that has been suggested in past studies of the pulsations in Be stars in such X-ray binaries is the phenomenon of period drifting - i.e. slowly varying periods, probably associated with Non-Radial Pulsations (NRP) from the Be star \citep{2002MNRAS.332..473C, 2005AJ....130.2220S}. Though the expectation for such period changes being associated with NRP behaviour does not comfortably fit with stellar oscillation models, it is extremely unlikely that such changes could be associated with a changing orbital period on the timescales of the OGLE data. \citet{2002MNRAS.332..473C} discussed this possibility in the context of a possible triple system for SXP323. Since aliased non-radial pulsations are believed to be responsible for the sinusoidal modulations in many OGLE light curves, we have investigated this phenomenon of a drifting period in several systems where the periodicity is strong enough to determine the period accurately on annual timescales. 

\begin{itemize}

\item SXP214 shows a clear and steady drift in period - the first time a drifting periodicity has been determined in this system (Figure~\ref{fig:drift214}). To determine the true drift rate we assume that the measured period of around 4.55d shown in the figure is actually the beat between the true period and the one day sampling of the OGLE data, but as stated previously, there is a degeneracy in this possible solution. Assuming a constant drift rate over the period of the observations, the degeneracy indicated in Figure~\ref{fig:aliasdegen} leads to two possible solutions for the true period and drift: either a period of 0.818d with an associated drift of +15.0s/yr or a period of 1.277d with a drift of  $-$37s/yr.
 
\begin{figure}
\includegraphics[scale=0.4]{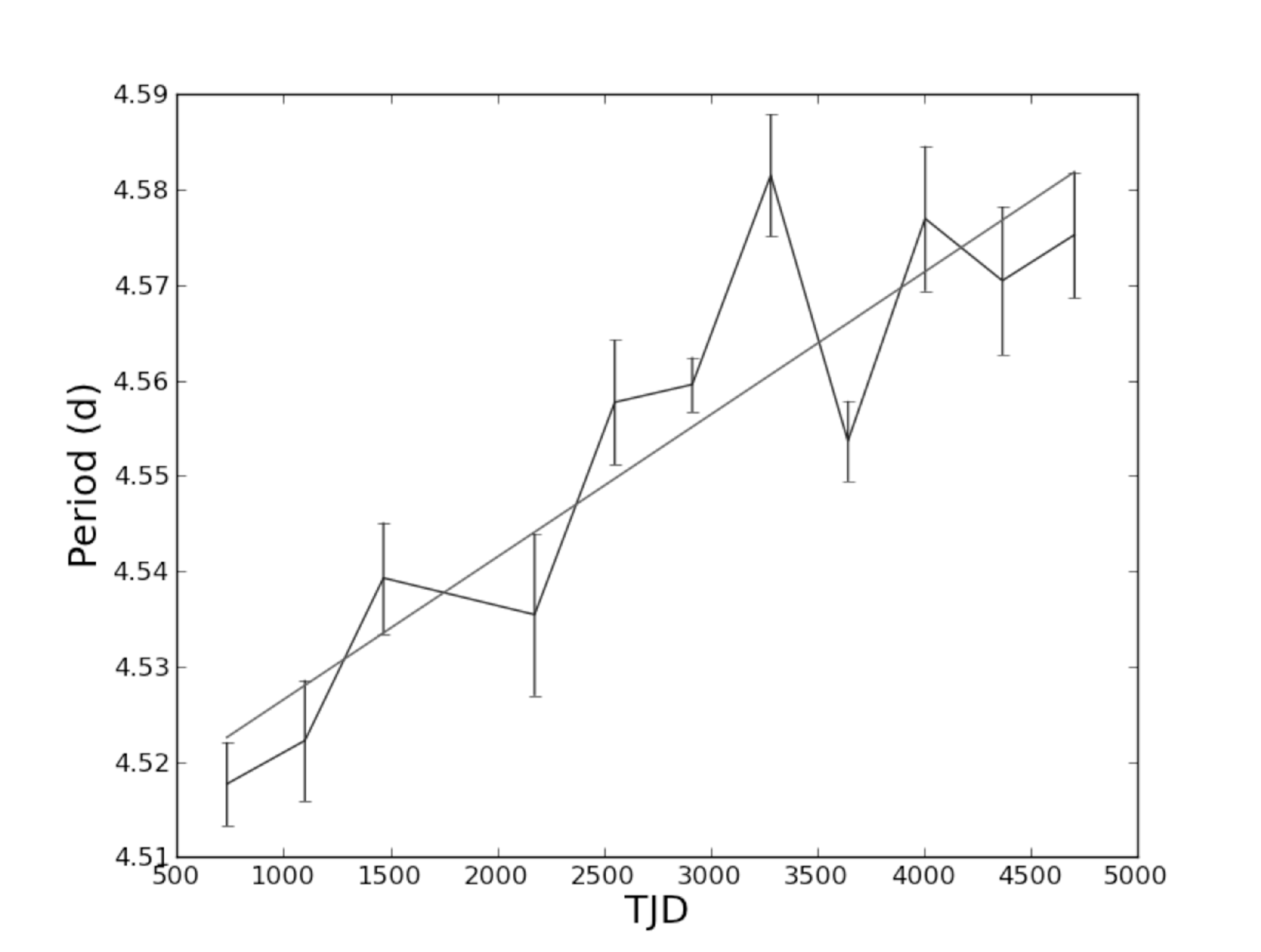}
\caption{Evidence of periodicity drift in SXP214.}
\label{fig:drift214}
\end{figure}

\item SXP323 shows a different behaviour (Figure~\ref{fig:drift323}). While there is still an overall drift in period, it appears there may not be a simple monotonic progression like that seen in SXP214. Nonetheless a simple linear fit to the data over the duration of the OGLE observations gives a change in the period of +4.92s/yr. Superimposed on this steady change is an apparent long term modulation with a period $\sim$2000d, but the limited length of the data set means that the evidence for such a modulation is tentative at best.

\begin{figure}
\includegraphics[scale=0.4]{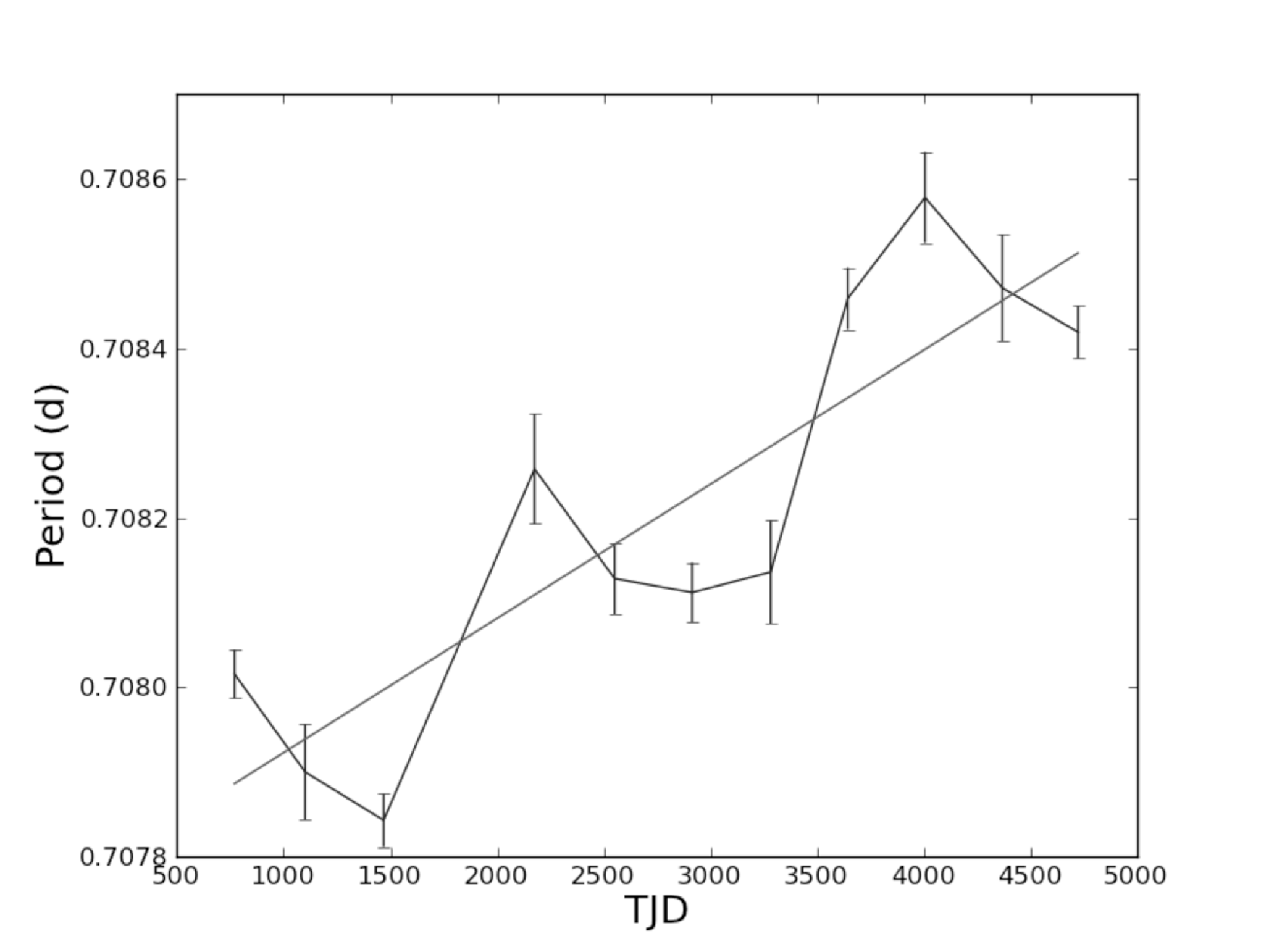}
\caption{Evidence of periodicity drift in SXP323.}
\label{fig:drift323}
\end{figure}

\item SXP701 also exhibits a convincing period drift over the duration of the OGLE observations. In this source there is evidence for two possible observed periods ($\sim$0.2845d and $\sim$0.282d) with the main power jumping from the former to the latter after two years. If a linear fit is made to the most persistent second period then a period drift of +5.26s/yr is obtained - similar to that seen in SXP323. We note that \citet{2005AJ....130.2220S} obtained -2.6s/yr for the period drift in this object from the shorter MACHO data set - the negative rate of change possibly arising from the misleading effect seen here in the OGLE data whereby different frequencies can take turns to dominate.

\end{itemize}

There has been some discussion in the literature as to whether the occasionally reported period drifts in Be stars (eg \citet{1996A&A...309..787S}) are real, or due to unresolved beating between similar periods with differently changing amplitudes. A discussion on this topic may be found in Baade (1999) but with no clear conclusions. Certainly the data presented here represent some of the longest time bases for such observations and further investigation into the OGLE archives for similar stars and periods may be profitable.

\subsection{Superorbital periods}

It has been suggested by \citet{2011MNRAS.413.1600R} that these systems may exhibit optical modulation on longer timescales than their orbital periods. Furthermore those authors present results which indicate a possible correlation between orbital and superorbital periods (see Figure 45 in their paper). Though their plot looks convincing, care must be taken when creating such data sets. Since the total length of the OGLE data is $\sim$5000d, then any period longer than half of that must be treated with some scepticism. This effectively removes 6 of the 19 objects from their plot at the longer period end.

At the shorter period end the work presented here casts doubt on the value for several periods quoted in \citet{2011MNRAS.413.1600R} as orbital. If the shape of the modulation is strongly sinusoidal then there is a very good chance that it arises as an alias of an underlying NRP. For example, the period of 36.4d given for SXP7.92 appears from our work to be a strong case for being the alias of an NRP, as is also the case for SXP101 (21.9d) and SXP3.34 (11.1d) - the latter which \citet{2011MNRAS.413.1600R} describe as a ``marginal" case anyway. If one removes these objects from their plot, together with the long period ones, then the case for a relationship between orbital and possible superorbital periods in the remaining 10 objects is greatly reduced. There is no doubt that Be stars in the SMC (and probably elsewhere) show considerable creativity in their modes of variability \citep{2002A&A...393..887M}, so great care must be taken when claiming actual periodicities when, instead, such observational changes may represent general long-term timescales for change in the mass outflow from such stars.
If such superorbital periods exist, then the warping of the circumstellar disk by an inclined neutron star orbital plane may provide an explanation \citep{2011arXiv1106.2591M}.

\subsection{The V-band light curves}

Although the V-band data coverage is only $\sim$10\% of the I-band data, we searched the V-band data to determine if any of the types of periodicities were more systematically visible in that band. A few periodicities were significantly detectable in the V-band lightcurves, namely those in SXP7.78, SXP293 and SXP1323; another, SXP327, is detected but below a formally significant level. It is difficult to explain why these particular sources exhibit periodicities in the V-band while others do not. While we might expect persistent pulsations of the donor star to be easier to detect in the V-band than any decretion disk modulation, both confirmed orbital and suspected pulsation periods are included in this very small subset of detected periods. Although the amplitude of these modulations is certainly higher than average, we cannot say for sure that large amplitude is the only reason for detection. Unfortunately, other sources that we might expect to detect on such a basis, such as SXP18.3 and SXP893, exhibit their modulation only transiently, and have only 16 and 4 V-band data points respectively in the relevant time periods. The data for SXP756 show only one periastron outburst, which is strong but obviously offers no basis for period detection. In all these cases then, we can state that the limited data coverage prohibits any possibility to detect the periodicities, and we should make no physical inference on the lack of detection in the V-band.

\section{Conclusions}

\begin{figure*}
\includegraphics[scale=0.68]{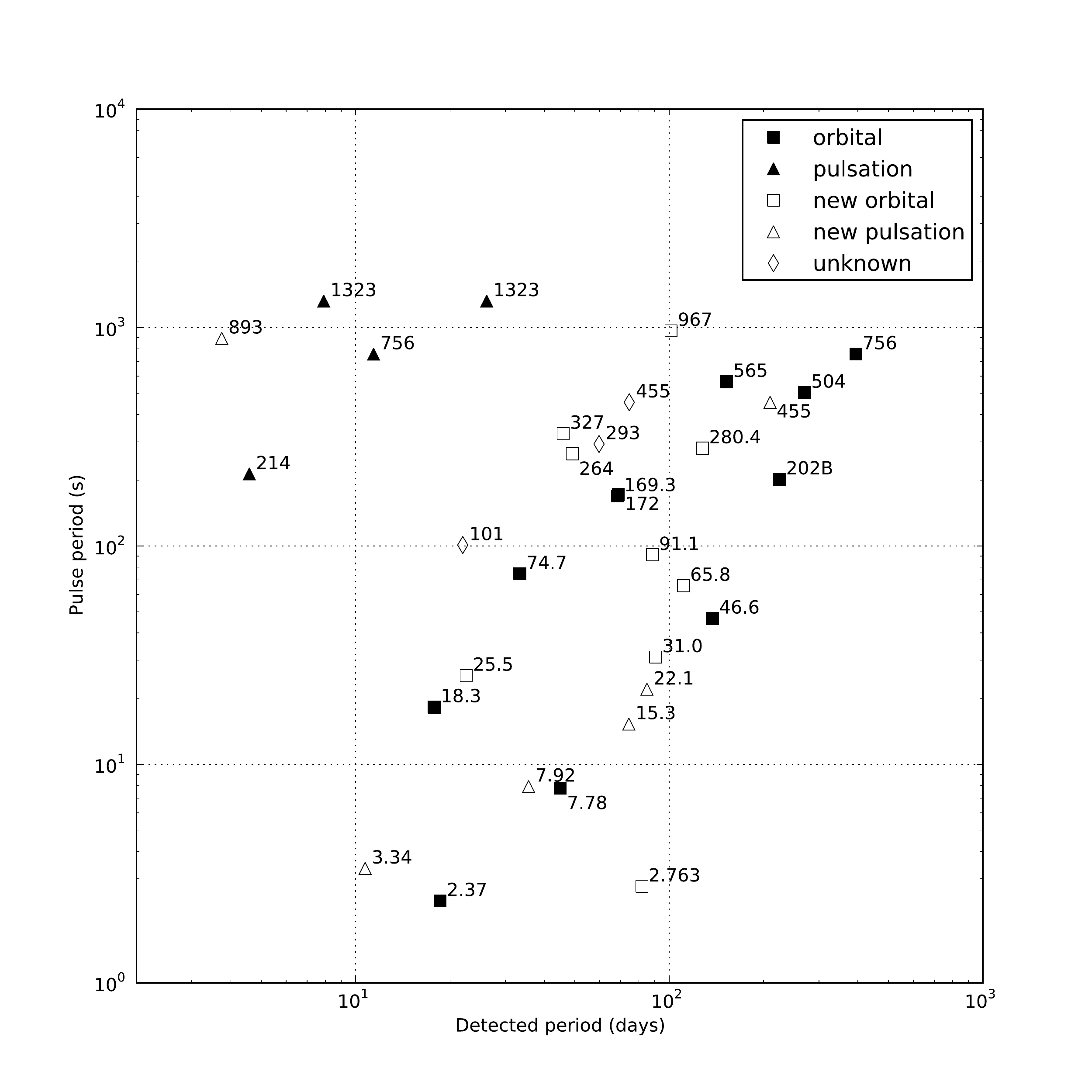}
\caption{A P$_{\rm{orb}}$-P$_{\rm{spin}}$ diagram constructed assuming all detected periodicities, coded to indicate the suspected mechanisms for the periodicities.}
\label{fig:cccorbet}
\end{figure*}

It is clear that the use of ground-based optical light curves like those in the OGLE-II and OGLE-III programs to determine the orbital periods of Be/X-ray binary systems can be an effective technique. But at the same time, it can be misleading, as other mechanisms can produce periodicities that, when combined with typical daily sampling, alias into the typical orbital period regime. This can be illustrated by use of a P$_{\rm{orb}}$-P$_{\rm{spin}}$ diagram (Figure~\ref{fig:cccorbet}) in which we plot all the periodic signals detected, regardless of their suspected origin. The systems for which the modulation is known, or strongly suspected, to be of orbital origin, lie in the traditional zone of the diagram associated with the Be/X-ray binaries. However, the systems for which the origins of the periodicity may be aliased NRPs (or other sinusoidal variations of unknown origin) are outliers to this distribution, and coincidentally fall in the region traditionally associated with supergiant systems. 

We present a method that allows the shape of the folded light curve to be used to distinguish between orbital periods and other mechanisms in at least some cases, but we point out that this is still not a guarantee. For now, the only unambiguous way of ascertaining the orbital periods of the Be/X-ray binaries comes from X-ray measurements, either by looking for periodic emission or using Doppler methods to fit the orbital parameters directly. 

\section*{Acknowledgments}

The OGLE project has received funding from the European Research Council under the European Community's Seventh Framework Programme(FP7/2007-2013) / ERC grant agreement no. 246678 to AU. VM's research is supported by the University of Cape Town and the National Research Foundation of South Africa.

\label{lastpage}

\end{document}